\documentclass[twocolumn]{aastex631}
\usepackage{graphicx}
\usepackage[dvipsnames]{xcolor}
\usepackage{float}
\usepackage{placeins}
\usepackage{lineno}

\newcommand{\ECF}{emission contribution function }
\newcommand{\ECFs}{contribution functions }


\begin{document}
\title{A grid of contribution functions to help the interpretation of time-resolved observations of brown dwarfs}

\author[0000-0002-0786-7307]{Myrla Phillippe}
\affiliation{Department of Physics, University of Central Florida, 
4111 Libra Dr, Orlando, FL 32816, USA, myrla.phillippe@ucf.edu}

\author[0000-0001-7356-6652]{Theodora Karalidi}
\affiliation{Department of Physics, University of Central Florida, 
4111 Libra Dr, Orlando, FL 32816, USA}

\author[0000-0003-0192-6887]{Elena Manjavacas}
\affiliation{AURA for the European Space Agency (ESA), ESA Office, Space Telescope Science Institute, 3700 San Martin Drive, Baltimore, MD,21218 USA}
\affiliation{Department of Physics and Astronomy, Johns Hopkins University, Baltimore, MD 21218, USA}

\author[0000-0001-5254-6740]{Natalia Oliveros Gomez}
\affiliation{Department of Physics and Astronomy, Johns Hopkins University, Baltimore, MD 21218, USA}

\author[0000-0003-3714-5855]{Daniel Apai}
\affiliation{University of Arizona, 933 North Cherry Avenue, Tucson, AZ 85721, USA}

\author[0009-0008-5805-880X]{Jonathan Fernandez}
\affiliation{Department of Physics, University of Central Florida, 4111 Libra Dr, Orlando, FL 32816, USA}

\author{Kieran M Manjrawala}
\affiliation{Department of Physics and Astronomy, Johns Hopkins University, Baltimore, MD 21218, USA}

\begin{abstract}
We present a grid of thermal emission contribution functions and their corresponding atmospheric spectra model for cloudy and cloud-free brown dwarfs, spanning effective temperatures  $T_\mathrm{eff}\in[500,2000]$[K], surface gravities $\log g=[3.5, 5.5]$ [cgs], sedimentation efficiencies $f_\mathrm{sed}=$1, 2, 3, 4 and 8, and eddy diffusion coefficients $\log K_\mathrm{zz}=[6,11]$. We use temperature-pressure profiles from the Sonora Diamondback models in conjunction with the \texttt{picaso} radiative transfer code and the cloud modeling code \texttt{virga}, to explore how different atmospheric parameters influence the vertical distribution of emitted radiation for atmospheres in chemical equilibrium. We confirm that cloud opacity significantly impacts atmospheric emission depth in prominent molecular features such as $CH_4$, $H_2O$, $CO$, $NH_3$, $CO_2$, and atomic features $Na$ and $K$. The atmospheric emission depth varies most notably in the $J-$band, and at mid-resolution spectra in the $Na$ and $K$ doublets. Our analysis highlights pressure-wavelength differences between cloudy and cloud-free atmospheres for characterizing L, L/T transition, and T-type brown dwarfs and demonstrates how contribution functions can guide targeted spectroscopic investigations. To facilitate observational planning, we provide an accessible database and web-based application that enable the community to compare observational data with our model grid of spectra and contribution functions with a wavelength range of  0.5 - 30 $\mu$m.

\end{abstract}


\section{Introduction} \label{sec:intro}
The comparison of theoretical models to brown dwarf and exoplanet observations enabled the characterization of these bodies, placing constraints on their temperature, gravity, and cloud properties. The L, L-to-T transition (hereafter L/T), and T spectral types define progressively cooler brown dwarfs, distinguished by changes in molecular absorption features and atmospheric variability { \citep{Marley2013,Helling2014}. L dwarf ($T_\mathrm{eff}\in[1300,2000]$K) spectra are shaped by strong $H{_2}O$ absorption bands, metal hydrides ($FeH$, $CrH$), and broadened alkali lines ($Na$, $K$, $Rb$, and $Cs$), while condensate clouds of Fe and silicates ($Al{_2}O{_3}$, $Mg{_2}SiO{_4}$, $MgSiO{_3}$) provide additional opacity and continuum reddening \citep{Cushing2005,Kirkpatrick2005,Marley2013}. 
T dwarf ($T_\mathrm{eff}\in[500,1300]$K) spectra exhibit prominent CH$_4$ absorption in the near-infrared together with strong $H{_2}O$ features and pressure-broadened alkali lines, and condensates such as sulfides and chlorides (e.g.,  $Na{_2}S$, $KCl$) may persist and contribute to cloud opacity \citep{Burgasser2006,Stephens2009,Morley2012}. Although most  brown dwarfs show some degree of rotational variability \citep[e.g.,][]{Metchev_2015,Buenzli2014}, such variability may remain undetectable, most notably due to limitations in observational sensitivity (e.g., smaller ground-based facilities compared to space-based observatories) or, in some cases, projection effects \citep{Vos2017, Vos2019}. In the L/T transition ($T_\mathrm{eff}\in[1100,1400]$K), the amplitude of rotational variability reaches its peak \citep[][and references therein]{, Radigan2014, Biller2017}}. Various mechanisms have been proposed to explain this behavior, including complex cloud dynamics  \citep{Tan2021}, thermal perturbations \citep{Robinson2014} and thermochemical instabilities \citep{Tremblin2016}. Cloud dynamics result in evolving, and possibly patchy, cloud cover that rotates in and out of view, producing flux changes \citep[e.g.,][]{Tan2021}. Observations strongly support this mechanism \citep[e.g.,][]{Radigan2012,Apai2013,Apai_2017,buenzli2012, Buenzli2014,faherty2014,Yang2015,Yang_2016,Lew_2016,manjavacas2019b, Suarez2023,suarez2023ultracooldwarfsobservedspitzer, Biller2024,Chen2025,oliverosgomez2025}. 
In the case of thermal perturbations, atmospheric temperature fluctuations (e.g., hot spots or waves) can cause variability even without cloud changes \citep{Robinson2014, Morley2014b}. 
However, \citet{Adams2025} showed that cloud-free thermal models are not able to match the high-amplitude, phase-consistent spectra of brown dwarf light curves.  
Lastly, \citep{Tremblin2016} argued that a $CO$–$CH_4$ fingering convective instability could cause temperature and compositional fluctuations, explaining spectral changes without clouds. While evidence for disequilibrium chemistry has been observed in a few targets, they are either colder Y dwarfs \citep[e.g.,][]{leggett2024}, or the disequilibrium chemistry appears confined to the upper layers of the atmosphere \citep[e.g.,][]{oliverosgomez2025} and is potentially connected with auroral heating \citep{McCarthy2024,Chen2025,nasedkin2025}. 
Thermal perturbation and thermochemical instabilities effects could possibly contribute secondarily as  mechanisms the to observed variability, however, atmospheric cloud variations appear to be the main driver of brown dwarf variability \citep[e.g.,][]{Radigan2012,Apai2013, Lew2020, Tan2021, Miles2023JWST,suarez2022,Suarez2023}
. 


Time-resolved observations with NIRSpec  \citep{Jakobsen_NIRSPEC}  and MIRI \citep{Wright2015_MIRI}  onboard the \textit{James Webb Space Telescope} (\textit{JWST}) \citep{Gardner2006JWST} provided the community with an unprecedented simultaneous wavelength coverage. 
Since different wavelengths trace different depths in an atmosphere, it is possible to measure the variation in opacity at different layers. 
This transforms the one-dimensional spectrum into a pressure–wavelength (2D) map and enables the identification of the depth at which different spectral features originate in an atmosphere \citep[e.g.,][]{Yang_2016,Lothringer2018, Chen2024}. Applied on time-resolved observations, this gives us a glimpse of the 3D structure of an atmosphere \citep[e.g.,][]{Chen2025, akhmetshyn2025}. \textit{JWST}  opened a new window in the study of atmospheres, expanding spectroscopic coverage into a broad, contiguous near- to mid-infrared regime. Prior observations were limited to narrow wavelength windows defined by individual instruments~-- including \textit{Spitzer} (\citealp{Werner2004}; used, e.g., in \citealp{Metchev_2015, Yang_2016}), \textit{HST} (\citealp{Williams1996}; used, e.g., in \citealp{buenzli2012, buenzli_2015, Yang_2016, Manjavacas_2019}), \textit{TESS} (\citealp{Ricker2015}; used, e.g., in \citealp{Apai2021}), and \textit{Kepler}/\textit{K2} (\citealp{Borucki2010, Howell2014}; used, e.g., in \citealp{Gizis2015, Metchev_2015})~-- and gave us access to a limited slice of these atmospheres. 
{Furthermore, the lack of temporal overlap between observations at different wavelengths made it impossible to constrain simultaneous atmospheric variations across different pressure levels, limiting our ability to characterize the 3D atmospheric structure.}

A few observations, like the \textit{Spitzer} Cycle-9 Exploration Science Program Extrasolar Storms (Program ID: 90063, PI: Apai), partially broke this limitation by combining \textit{Spitzer} and \textit{HST} observations in an effort to study how clouds change as a function of Spectral Type (SpT) and pressure in the atmosphere. 
Extrasolar Storms allowed the community to get a first \textit{near simultaneous} glimpse of different pressure layers for 4 targets (two L5 targets: 2MASS J15074769-1627386 and 2MASS J18212815+1414010; a T2 target: SIMP J013656.5+093347.3; and  a T6 target: 2MASS J22282889-4310262) over two epochs \citep{Yang_2016, Apai_2017}. However, to date, the full informational content of these observations has not yet been leveraged by the community at large. JWST now gives us a \textit{simultaneous} glimpse of multiple layers with the same instrument.

Here, we aim to facilitate the characterization of archival observations and the planning of feature observations by presenting a grid of emission contribution functions for atmospheres spanning the full L- to T- spectral types.
Emission contribution functions offer an approximation of the pressure layer that each wavelength is sensitive to, and are a useful tool for studying the vertical distribution of emitted light in brown dwarf atmospheres \citep[e.g.,][]{buenzli2012,Yang_2016, Vos_2023, McCarthy2024, Chen2024}. Recent work by \cite{Chen2024} compared the \ECF of a cloud-free brown dwarf atmosphere to that of a cloudy brown dwarf atmosphere and found that the addition of clouds significantly altered the contribution function. 
They emphasized the thermal emission contribution functions as useful guides for understanding atmospheric probing. 
However, while \ECFs have been acknowledged as useful guides for 
determining how different atmospheric layers contribute to the observed thermal emission, the community lacks a comprehensive \ECF source and to date \ECFs are custom-made on a case-by-case. We address this need by presenting a grid of 2407 cloudy and 80 cloud-free brown dwarf contribution functions and the corresponding emission spectra for comparison against observations. Our models cover the parameter space of $T_\mathrm{eff}\in[500,2000]$K (with a step of 100 K), $\log g\in[3.5,5.5]$ [cgs] (with a step of 0.5), sedimentation efficiency parameter $f_\mathrm{sed}\in[1,2,3,4,8]$ and a constant with-altitude eddy diffusion coefficient $\log K_\mathrm{zz}\in [6,11]$. Our models have a constant, solar metallicity and C/O abundance. Our work aims to maximize the scientific return of both upcoming observations and archival data, as it allows users to constrain the average pressures probed by their (archival or planned) observations and constrain the 2D structure of their target atmospheres.



Here, we study trends of our contribution functions as a function of different model parameters (T$_\mathrm{eff}$, $\log g$, $f_\mathrm{sed}$ and $K_\mathrm{zz}$). These trends can help inform observational strategies by highlighting the specific spectral features they need to target depending on the purpose of the observations. Our models are available on Zenodo \footnote{ doi: 10.5281/zenodo.17834338} and conveniently accessible through a Graphical User Interface (GUI) that allows users to query individual spectra or \ECF outputs without downloading the full database\footnote{https://ltcfgrid.research.ucf.edu/}. This tool was created to facilitate model comparisons with observations and assist observational planning (see section~\ref{sec:discussions}). 

{
In Sect.~\ref{sec:style} we give an overview of the process used to derive the average pressures probed in our model atmospheres (the \ECF). In Sect.~\ref{sec:results}, we present our results and comparisons of the pressures probed focusing on 
the $J-$, $H-$ and $K-$ bands, as well as $CH_4$, $H_2O$, $CO$, $NH_3$, $CO_2$, $Na$ and $K$ doublets. In Sect.~\ref{sec:discussions}, we discuss the application of our grid to archival \textit{HST} and \textit{JWST} observations, discuss our results focusing on the physical implications of $K_\mathrm{zz}$ for the cloud formation, and provide details about the GUI we designed to facilitate community access to our grid. Finally, in Sect.~\ref{sec:conclusion} we present our conclusions and discuss future updates possible for our grid.}

\section{Methods} \label{sec:style}

\subsection{Thermal Profiles and Cloud Profiles}
We used the temperature--pressure (TP) profiles from the Sonora Bobcat \citep[][data available on Zenodo: \citealt{marley_bobcat_zenodo}]{Marley_2021} and Diamondback \citep[][data available on Zenodo: \citealt{morley_diamondback_zenodo}]{Morley2024} grid models. The Sonora Bobcat models focus on cloud-free, chemical equilibrium atmospheres for atmospheres with $T_\mathrm{eff}\in[200,2400]$K. The Sonora Diamondback models extend this framework by incorporating clouds, and include updated molecular opacities (including $Na$, $K$, $TiO$, $VO$, $FeH$, and $CH_4$), and a finer TP grid (1460 points vs. Bobcat’s 1060). These improvements allow Diamondback to better capture the spectral behavior of cloud-dominated L and early T-type atmospheres, producing notably redder near-infrared spectra under equivalent conditions. The Diamondback models focus on atmospheres hotter than $T_\mathrm{eff}=900$K as below this limit the evolutionary models converge with the Bobcat models (see the discussion of the Sonora Diamondback grid in \citealt{Morley2024}). 

\begin{figure*}[htbp!]
    \centering
    \includegraphics[width=1\textwidth]{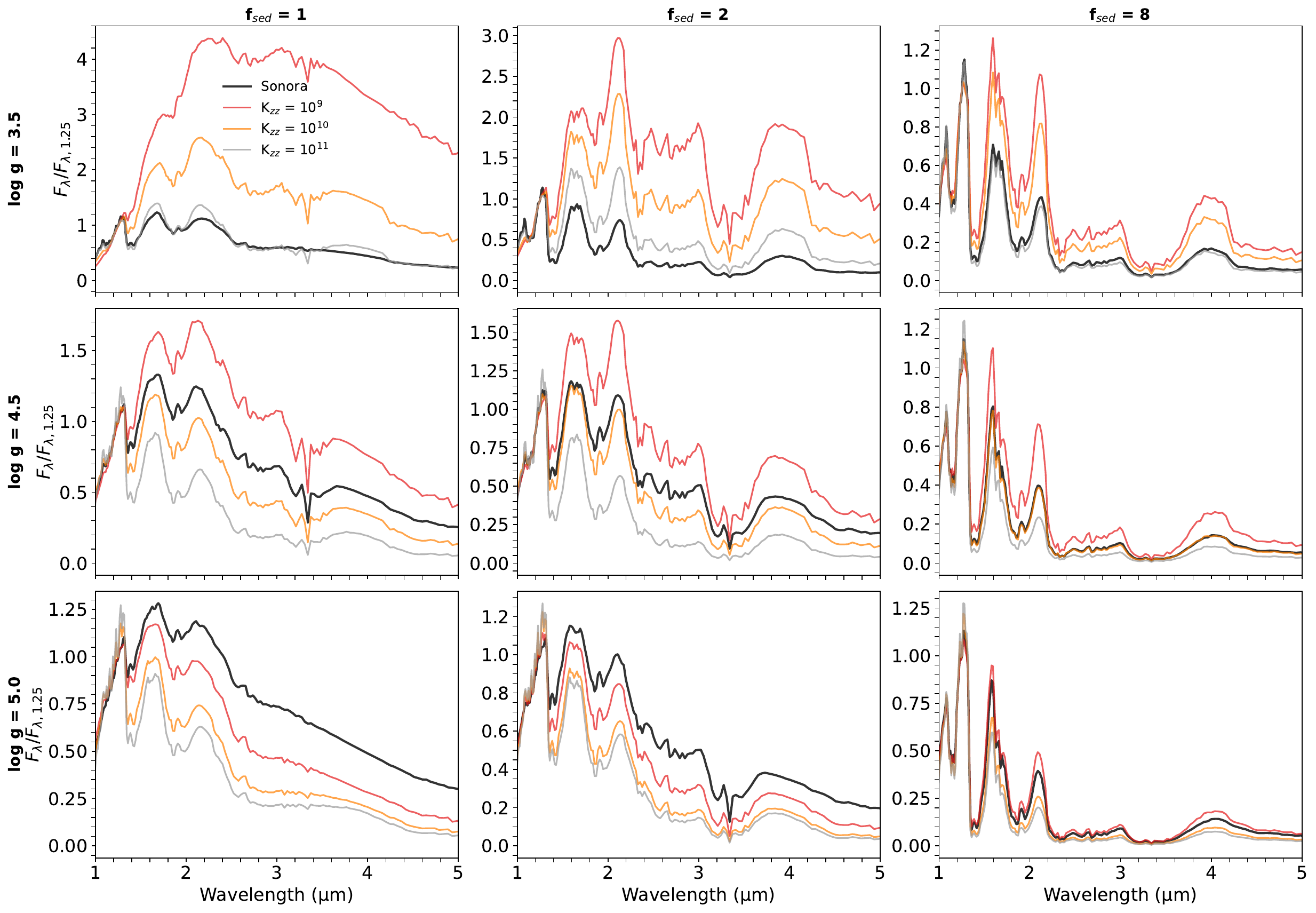}
    \caption{Model spectra for atmospheres at $T_\mathrm{eff} = 1200\,\mathrm{K}$ with  $\log g =3.5$ (top panels), 4.5 (middle panels) and 5.5 (right panels) and $f_\mathrm{sed}=1$ (left column), 2 (middle column) and 8 (right column). The atmospheres have $\log K_\mathrm{zz}=$9 (red lines), 10 (orange lines) or 11 (gray lines). The corresponding Sonora Diamondback models that used a altitude-dependent $\log K_\mathrm{zz}$ are also shown (black lines) for comparison.  }
  \label{fig:DBSpecComp}
\end{figure*}

In Fig.~\ref{fig:DBSpecComp}  we 
compare our grid model spectra against the Sonora Diamondback spectra \citep{morley_diamondback_zenodo}  at $T_\mathrm{eff} = 1200\,\mathrm{K}$  over the 1--5\,$\mu$m wavelength range. The spectra are normalized to their flux at $1.25\,\mu\mathrm{m}$. The model atmospheres have $\log g =3.5$ (top panels), 4.5 (middle panels), and 5.5 (bottom panels) and $f_\mathrm{sed}=1$ (left column), 2 (middle column),  and 8 (right column). Finally, the models have a constant-with-altitude $\log K_\mathrm{zz} =9$ (red lines), 10 (orange lines) and 11 (gray lines). Also shown are the corresponding Sonora Diamondback spectra (black lines) that use an altitude-dependent $K_\mathrm{zz}$. Our spectra are comparable with the Sonora Diamondback ones at low gravities ($\log g = 3.5$)  with higher $K_\mathrm{zz}$ (e.g., for $f_\mathrm{sed}=1$ we get $\chi^2_\mathrm{red} = 0.012$ at $\log K_\mathrm{zz}=11$ vs $\chi^2_\mathrm{red} = 11.27$ at $\log K_\mathrm{zz} = 9$)  and at high gravities ($\log g = 5.0$) with lower $K_\mathrm{zz}$. Lastly, we note that the higher $f_\mathrm{sed}$ is (i.e., the closer the atmosphere is to a cloud-free model), the better the fit of the Sonora Diamondback spectra to our $\log K_\mathrm{zz}=11$ model is.  For example,  at $\log K_\mathrm{zz}=11$,  $\log g = 3.5$ and $f_\mathrm{sed}=8$ we get  $\chi^2_\mathrm{red} = 2.76 \times 10^{-3}$.   

We generated contribution functions using \texttt{picaso} v3.2 \citep[][code available on Zenodo: \citealt{batalha_picaso_zenodo}]{Mukherjee2023} and the TP input from Sonora Bobcat and Diamondback. {
{For our atmospheric opacities, we created a custom-made database using data from the line lists of \citealt{lupu2022} downsampled at $R=500,000$. We then again down-sampled our spectra at R $\sim$5,000,} using the \texttt{mean\_regrid} function in \textsc{picaso} \citep{Mukherjee2023}, which rebins a high-resolution spectrum to a target spectral resolution by computing the mean flux within each spectral bin. 
The resolution of $R \sim 5{,}000$ was chosen to exceed the spectral 
resolution of all instruments currently used to observe brown dwarfs. 
For space-based telescopes for example, these include the \textit{JWST} NIRSpec high-resolution gratings 
($R \sim 2{,}700$; \citealt{Jakobsen_NIRSPEC}; e.g.\ \citealt{Faherty2024, 
Lew2024, Manjavacas2024, Gandhi2023}), the \textit{JWST} NIRSpec medium-resolution 
gratings ($R \sim 1{,}000$; \citealt{Jakobsen_NIRSPEC}; e.g.\ 
\citealt{Miles2023JWST}), and the \textit{JWST} low-resolution NIRSpec PRISM/CLEAR ($R \sim 30$--$300$; 
\citealt{Jakobsen_NIRSPEC}) and MIRI LRS ($R \sim 40$--$160$; 
\citealt{Wright2015_MIRI}), as well as the \textit{HST} WFC3/G141 
($R \sim 130$; \citealt{Manjavacas_2019}). This ensures that our grid 
spectra can be straightforwardly downsampled to match any of these 
instruments for direct comparison with observations. While ground-based high-resolution spectrographs such as CRIRES+ ($R \sim 100{,}000$; \citealt{deregt2025}) or IGRINS ($R\sim45,000$; \citealp{yuk2010,mace2016}) exceed our chosen resolution, such instruments have been used for narrow band variability studies aiming to Doppler image a brown dwarf in a single band/pressure layer \citealp{Crossfield2014, Chen2024, ishikawa2025}. The primary science driver for our grid is multi-wavelength time-resolved observations, for which our chosen $R \sim 5{,}000$ provides sufficient resolution for existing and future observations. Our grid spans a wide wavelength range at sufficient resolution to capture key molecular and atomic features and probe atmospheric variability across a broad pressure range in brown dwarf atmospheres. 

Our database included  $CH_4$, $CO$, $CO_2$, $FeH$, $H_2$, $H_2O$,  $H_2S$, K , $N_2$, $NH_3$, $Na$, $PH_3$, $TiO$ and $VO$. Our model atmospheres are in chemical equilibrium, and we assumed solar C/O ratio and metallicity.

Cloud condensation follows the \cite{AckermanMarly2001} model as implemented in the \texttt{Virga} code v0.0 \citep[][code available on Zenodo: \citealt{batalha_virga_zenodo}]{batalha2025}.
Depending on the model $T_\mathrm{eff}$, our models include clouds composed of  $KCl$, $ZnS$, $Na_2S$, $MnS$, $Cr$, $MgSiO{_3}$, $Mg{_2}SiO{_4}$, $Al{_2}O{_3}$, and $Fe$; clouds included in the work of \citep{Morley2024}. }
  At  $T_\mathrm{eff}\in [500, 600]$K we include clouds of  $KCl$, $ZnS$, $Na_2S$, $MnS$,  and $Cr$.  For  $T_\mathrm{eff}\in [700, 800]$K we added clouds of \text{MgSiO\textsubscript{3}} and for $T_\mathrm{eff}\in [900, 2000]$K we include $MgSiO{_3}$, $Mg{_2}SiO{_4}$, $Al{_2}O{_3}$, and $Fe$. Cloud condensation of the species included occurs following the methods of \cite{batalha2025}.

\subsection{$f_\mathrm{sed}$ \& $K_\mathrm{zz}$} \label{sect:methods_fsedkzz}
The sedimentation efficiency parameter $f_\mathrm{sed}$ \citep{AckermanMarly2001} dictates the cloud particle sizes and controls the vertical extent of clouds. Smaller $f_\mathrm{sed}$ produces clouds that are optically thick, vertically extended, and composed of smaller particles. Larger $f_\mathrm{sed}$
produces optically thin clouds composed of larger particles.
Here, we followed the Sonora Diamondback grid and modeled atmospheres with $f_\mathrm{sed}$=1, 2, 3, 4 or 8, as well as cloud-free atmospheres.

{
The eddy diffusion coefficient ($K_\mathrm{zz}$) determines how efficiently the atmosphere transports material due to convective mixing within the atmosphere. Higher $K_\mathrm{zz}$ corresponds to more vigorous vertical mixing, efficiently transporting material from higher to lower atmospheric pressures; higher in the atmosphere.} Our model atmospheres have  $K_\mathrm{zz} \in[10^6,10^{11}]$ [cm$^2$s$^{-1}$].
While our grid includes $K_\mathrm{zz}$ values down to $10^6$ (relatively low vertical mixing ), some combinations (notably with $f_{\rm sed}=1$ or $2$) fail to show key absorption features producing a black-body-like spectrum or only show absorption features at much higher resolutions than the one used here (see also Sect.~\ref{sec:discussions}). Similarly, for models with high $K_\mathrm{zz}$ ($\in [10^{11},10^{12}]~\mathrm{cm}^2\,\mathrm{s}^{-1}$), the pressures probed by the cloudy models can approach those of the cloud-free profiles in some cases, particularly at large $f_{\mathrm{sed}}$ where the cloud opacity is already thin, as the mixing is too vigorous to allow a cloud to form. In this paper, we discuss some of these high-$K_\mathrm{zz}$ cases (up to $\log K_\mathrm{zz} = 12$); in the released grid however, we exclude all models where $K_\mathrm{zz}$ is so high that the resulting spectrum and \ECF resembles that of a clear atmosphere. 
It is important to note that while the Sonora Diamondback models \citep{Morley2024} used an altitude-dependent $K_\mathrm{zz}$, in this study, we use a constant-with-altitude $K_\mathrm{zz}$.
To ascertain the theoretical soundness of our $K_\mathrm{zz}$ choices, we calculated a plausible range of $K_\mathrm{zz}$ values. These have been determined using the radiative–zone lower limit \citep{moses2021chemical} and the convective–zone upper limit \citep{zahnle2014methane} as our lower $K_\mathrm{zz}^{\rm min}$ and upper $K_\mathrm{zz}^{\rm max}$ limits respectively. In particular, we defined $K_\mathrm{zz}^{\rm min}$ as: 

\begin{equation}\label{eq:4}
K_\mathrm{zz}^{\rm min} = 5 \times 10^8 \;\frac{H}{620\ \mathrm{km}}\;P_{\rm bar}^{-1/2}\left(\frac{T_{\rm eff}}{1450\ \mathrm K}\right)^4
\end{equation}

where $P$\textsubscript{bar} is the pressure at the top of the atmosphere, $H$ is the scale height and $T$\textsubscript{eff} is the effective temperature of the atmosphere. We defined $K_\mathrm{zz}^{\rm max}$ as:  

\begin{equation}\label{eq:5}
  K_\mathrm{zz}^{\rm max} = 2.5 \times 10^{10}
    \left(\frac{T_{\rm eff}}{600\ \mathrm K}\right)^{8/3}
    \left(\frac{g}{1000\ \mathrm{cm\,s^{-2}}}\right)
\end{equation}

\begin{figure}[htbp!]
  \centering
  \includegraphics[width= .52\textwidth]{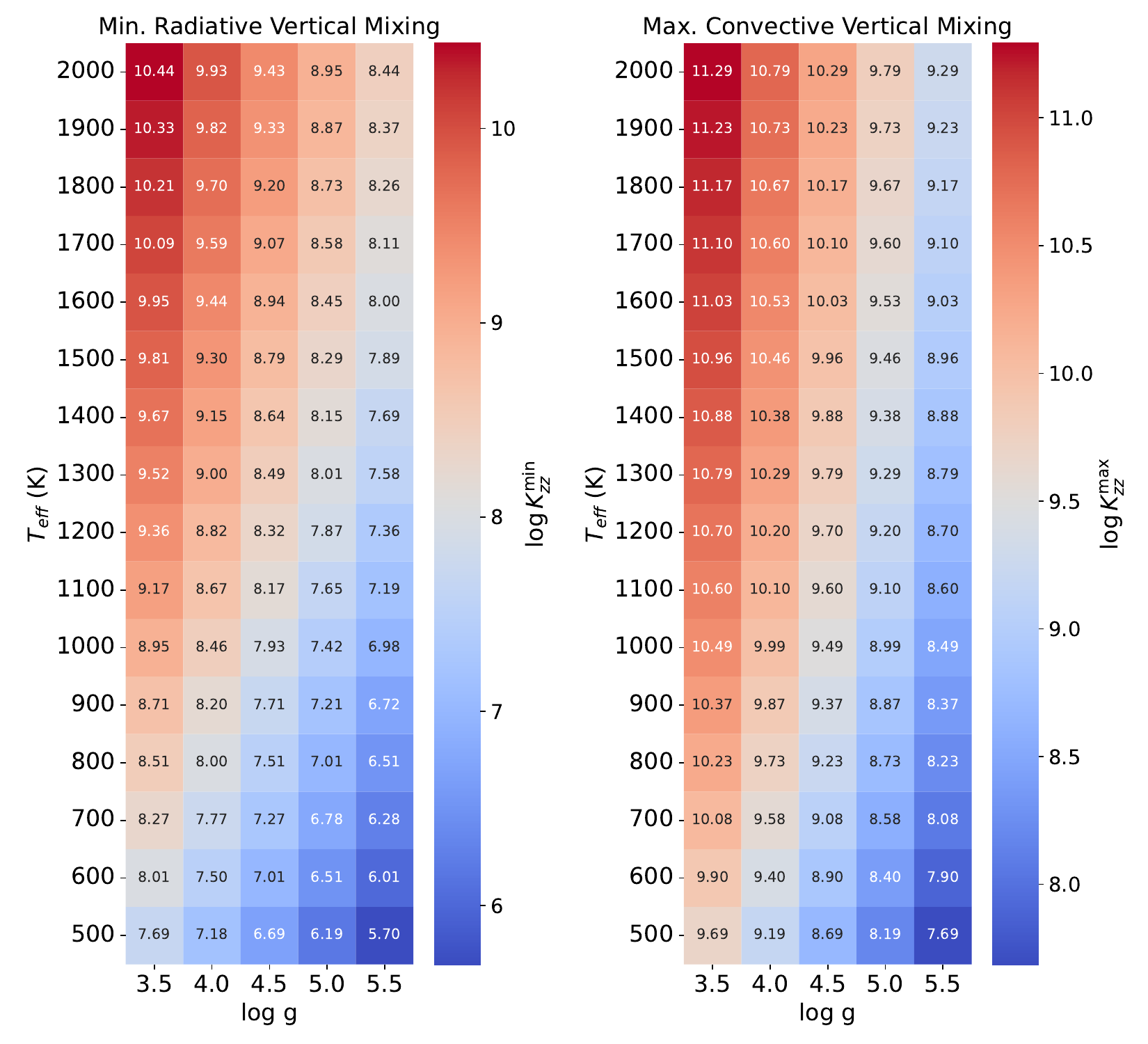}
  \caption{ Theoretical limits of $K_\mathrm{zz}$ defined here by the minimum expected $K_\mathrm{zz}$ of the radiative zone at the top of the atmosphere Eq.~\ref{eq:4} and the maximum of the convective zone Eq.~\ref{eq:5}. }
  \label{fig:Kzz_min_max}
\end{figure}

Considering these theoretical limits, users of our grid are advised to calculate and evaluate their own upper and lower limit bounds or alternatively use the bounds presented here  (Fig.~\ref{fig:Kzz_min_max})  when selecting models for comparison.  

\subsection{Average Pressures Probed}
We calculate the \ECF with \texttt{picaso} \citep{Mukherjee2022}, which 
defines the pressure probed at each wavelength as:

\begin{equation}\label{eq:1}
ECF(p, \lambda) = B_{\lambda} * e^{-\tau_{p,\lambda}} \frac{d\tau_{p,\lambda}}{dp}
\end{equation}

following  \cite{Lothringer2018}. 
Our \ECFs (cloudy and cloud-free) cover the parameter space presented in Table~\ref{Tabl:table1}. 

\noindent\textit{Cloud-free models:} \ECFs were generated using Sonora Diamondback's \citep{Morley2024} cloud-free TP profiles for $T_\mathrm{eff}\in[900,2000]$K. For $T_\mathrm{eff}\in[500,900)$K we used the Sonora Bobcat \citep{Marley_2021} TP profiles. Our models have $\log g$ of 3.5 to 5.5 [cgs]. For all models we used solar metallicity and a solar C/O ratio  \cite{lodders2009abundances}. \\
\textit{Cloudy models:} For the cloudy \ECFs we used the Diamondback TP profiles for $T_\mathrm{eff}\in[900,2000]$K, and the Bobcat TP profiles for $T_\mathrm{eff}\in[500,900)$K. Due to the constant with pressure $K_\mathrm{zz}$ we
used our grid spectra align most closely with the Diamondback spectra (calculated with a variable with altitude $K_\mathrm{zz}$) when $T_\mathrm{eff}$  and $K_\mathrm{zz}$ are both low or both high. 

\begin{figure*}[t]
  \centering

  \includegraphics[width=\textwidth]{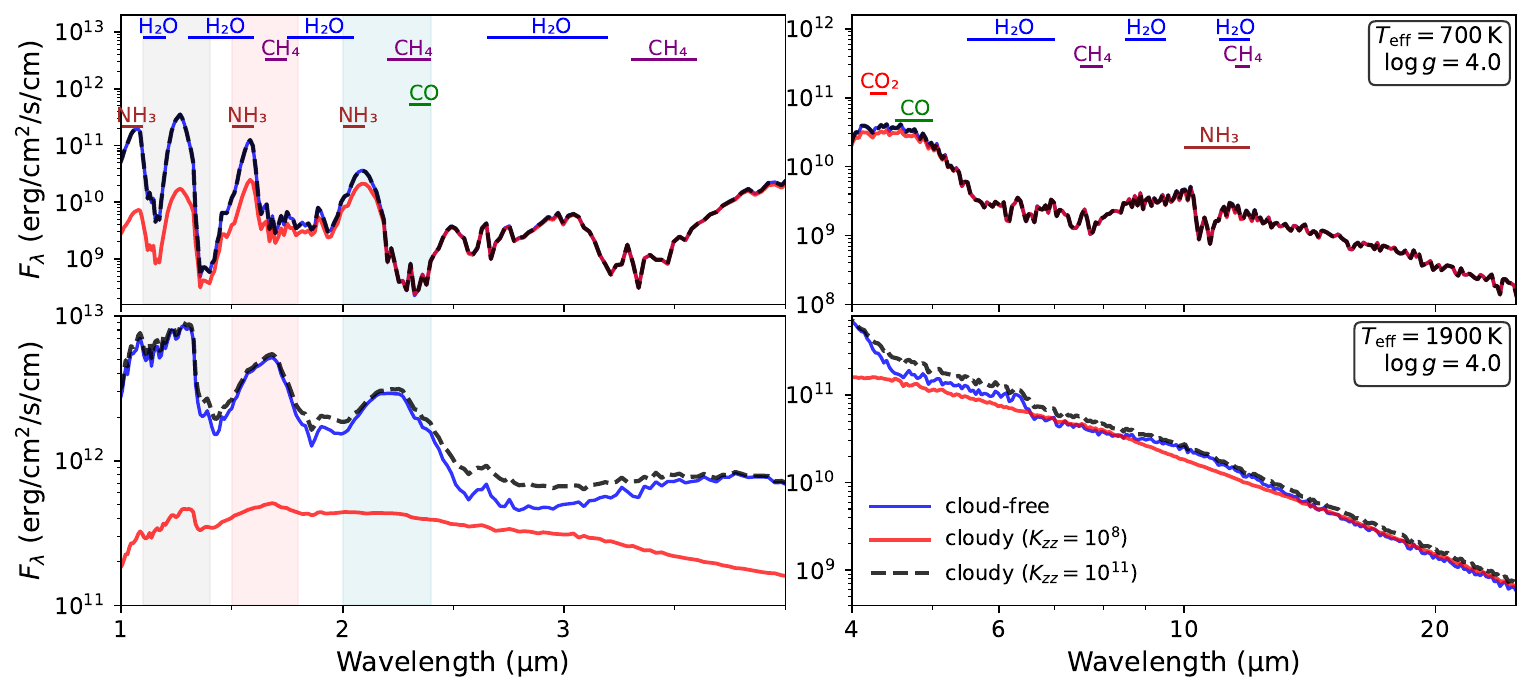}

  \vspace{0.02\textheight}

  \includegraphics[width=\textwidth]{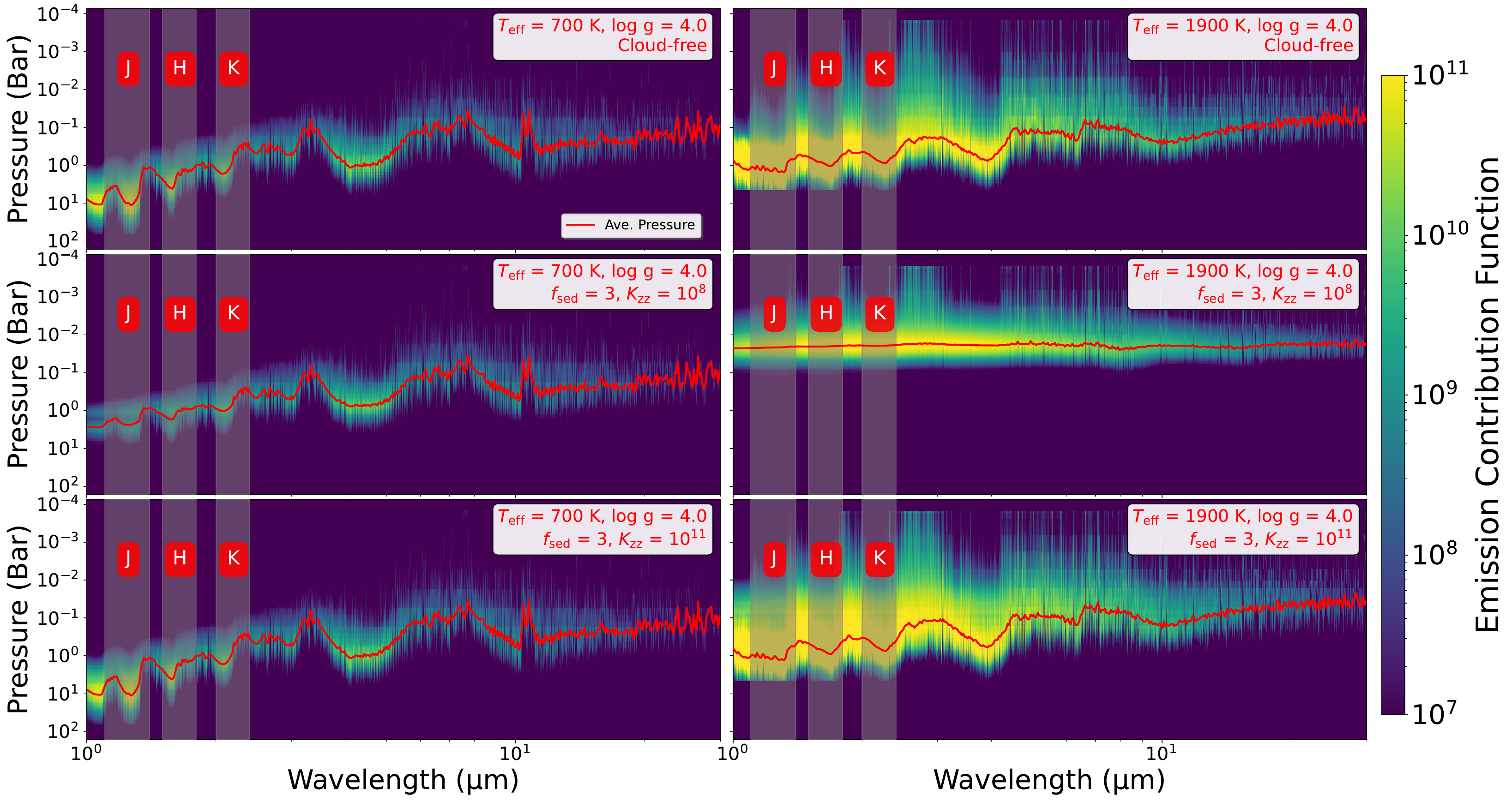}

  \caption{Top panels: Model spectra for cloudy (red and black lines) 
      and cloud-free (blue lines) atmospheres with $\log g=4.0$. Contribution functions of our atmospheres have $T_\mathrm{eff}=700$ K (top row) or 1900 K (bottom row). Highlighted are a few key absorption bands as well as the $J-$ (grey), $H-$ (pink), and $K-$ (blue) band. Bottom panels: Contribution functions for the clear (top row) and cloudy (bottom rows) atmospheres. The cloudy atmospheres have $f_\mathrm{sed}=3$ and $\log K_\mathrm{zz}=$8 (middle row ) or 11 (last row ). The higher $K_\mathrm{zz}$ for both the 700 K and 1900 K models results in the pressures probed being comparable to the clear atmosphere model, {while the lower $ K_\mathrm{zz}$ at $T_\mathrm{eff}=1900$ K produces a flattened \ECF as high altitude condensed clouds block the light from the deeper layers.}
  }
  \label{fig:side_by_side_cont}
\end{figure*}

In Fig.~\ref{fig:side_by_side_cont} we show example spectra ( top panels) and corresponding \ECFs for cloud-free (bottom panels; first row) and cloudy (bottom panels; second and third rows) atmospheres with $T_\mathrm{eff}=700$K (left panels) or $1900$K (right panels) and $\log g=$4.0. The clear atmosphere models probe the deepest pressures, 
particularly in the $J-$band [e.g., 12.7 (1.9) bars in the $J-$ ($K-$) band for the $700$ K model and 1.9 (1) bars for the $1900$ K model]. The moderately cloudy ($f_\mathrm{sed}=3$) $700$ K atmosphere model ($\log K_\mathrm{zz} =8$) probes lower pressures, up to an order of magnitude lower than the clear atmosphere models 
[e.g., 2.5 (1.1) bars in the $J-$ ($K-$) band]. On the other hand, the moderately cloudy $1900$ K ($\log K_\mathrm{zz}=11$) model probed comparable pressures to the clear model, due to the high atmospheric mixing combined with the lower sedimentation efficiency of this atmosphere. This is also supported by the maximum vertical mixing of $\log K_\mathrm{zz}=11$ shown in Fig.~\ref{fig:Kzz_min_max}. This highlights the need for the user to pay attention to the $K_\mathrm{zz}-f_\mathrm{sed}$ combinations they use. To produce the broader $\sim$9~$\mu$m silicate feature detected in objects such as VHS~1256b \citep{Miles2023JWST} and 
TWA~27b \citep{Patapis2025}, grain sizes $\lesssim 1~\mu$m are required in the upper atmosphere \citep{Luna2021}. \citet{Luna2021} show that models incorporating such small particles better reproduce the observed silicate feature than the \citet{AckermanMarly2001} cloud model. Since our grid is built on the \citet{AckermanMarly2001} sedimentation framework as implemented in \textsc{virga} 
\citep{batalha2025}, our models are not expected to reproduce a 
prominent silicate feature in the emergent spectrum. This is also discussed in \citet{Morley2024} where they find the same limitation in their Sonora Diamondback 
grid, which was also built on the \citet{AckermanMarly2001} framework.


\begin{table}[h!]
    \centering
    \begin{tabular}{ll}
        \hline
        \multicolumn{2}{c}{\textbf{Input Thermal Profile Parameters}} \\
        \hline
        Parameter & Range / Value \\
        \hline
        T$_\mathrm{eff}$  (Bobcat) & 500–800 K ( 100 K intervals) \\
        T$_\mathrm{eff}$  (Diamondback) & 900 - 2000 K ( 100 K intervals) \\
        Surface gravity ($\log g$) & 3.5–5.5 cgs (0.5 cgs intervals) \\
        $\log K_\mathrm{zz}$* & 6–11 cm$^2$ s$^{-1}$ \\
        $f_\mathrm{sed}$ & 1, 2, 3, 4, 8, no clouds \\
        C/O & Solar \\
        Metallicity ([M/H]) & Solar \\
        \hline
    \end{tabular}
    \caption{Summary of the parameter space covered in our grid. Our $T\mathrm{eff}$, $f\mathrm{sed}$, C/O, and M/H  follow the parameter space of the Sonora Bobcat \citep{Marley_2021} and Diamondback grids \citep{Morley2024}. \\ $^*$Note that following our findings in Sect.~\ref{sect:methods_fsedkzz} in our tool (see Sect.~\ref{sec:tool}), we include $\log K_\mathrm{zz}=6$ and 7 exclusively for $\log g \gtrsim 4.5$. Additionally, we include only the $\log K_\mathrm{zz}=11$ models that do not appear to be cloud-free.}
    \label{Tabl:table1}
\end{table}

\section{Results} \label{sec:results}

Here, we analyze the trends in the pressure probed in our model atmospheres for a few key wavelength bands: $J-$ (1.1--1.4 $\mu$m), $H-$ (1.5--1.8 $\mu$m) and $K-$ (2.0--2.4 $\mu$m) bands and a few select molecular and atomic features; $CH_4$, $H_2O$, $CO$, $NH_3$, $CO_2$ bands, K doublets, and Na doublet as a function of $T_\mathrm{eff}, \log g$, $f_\mathrm{sed}$ and $K_\mathrm{zz}$. As a reminder, our model atmospheres are in chemical equilibrium.

\subsection{The effect of gravity} \label{sec:grav}

\begin{figure*}[htbp]
  \centering
  \includegraphics[width=0.9\textwidth]{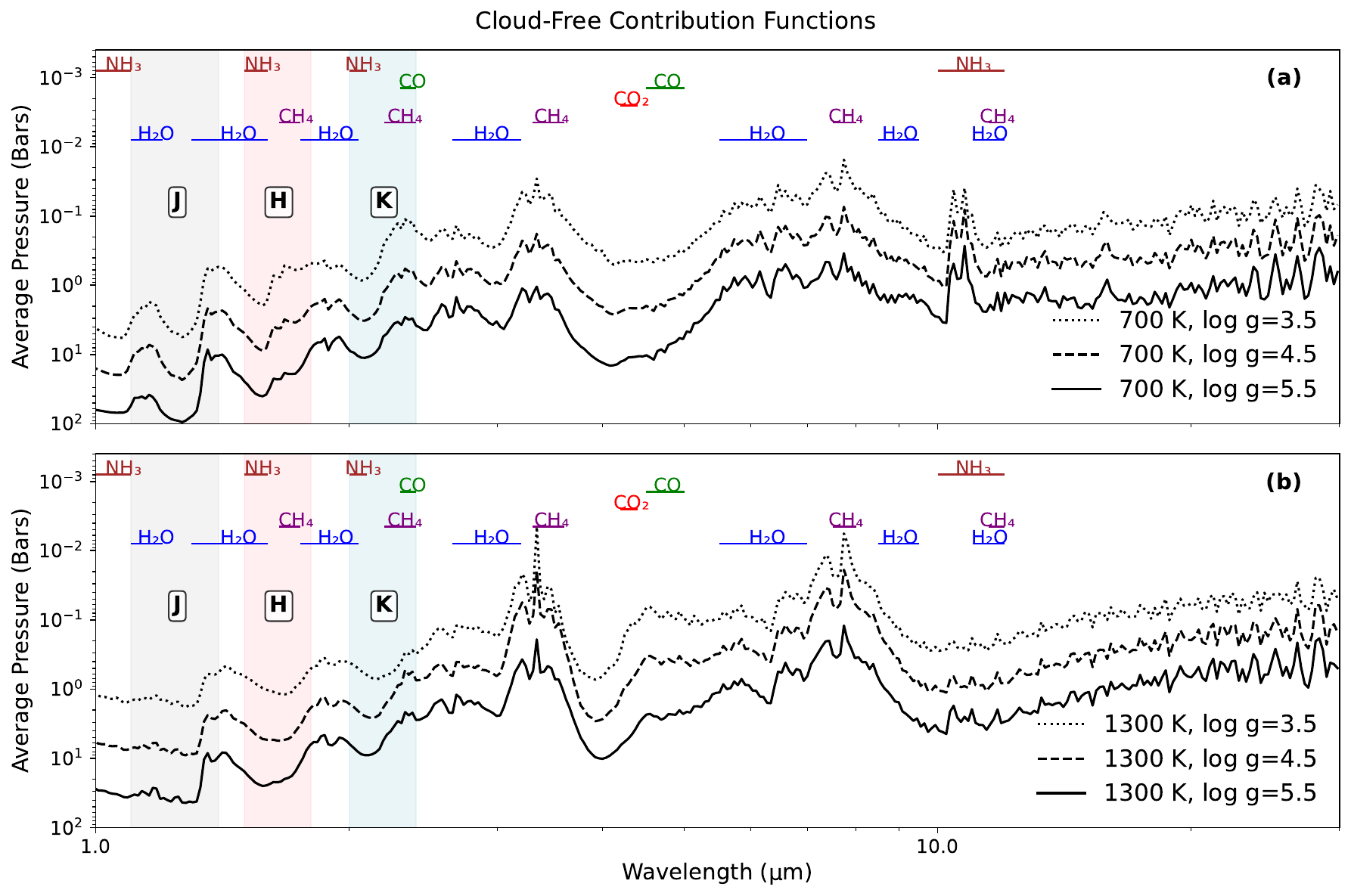}
  
  \caption{Contribution function for cloud-free grid models at 700 K (top) and 1300 K (bottom) at $\log g$ of 3.5 (dotted line), 4.5 (dashed line), and 5.5 (solid line). J-band , H-band, and K-band regions are highlighted in grey, pink and blue, respectively. We highlight some key absorption bands for H\textsubscript{2}O , CO\textsubscript{2} CH\textsubscript{4}  , 
  NH\textsubscript{3}, and CO.  It is evident here that deeper layers are probed at lower temperatures and higher gravity  in the J, H, and K bands.}
  \label{fig:cfree}
\end{figure*}

In Fig.~\ref{fig:cfree}  we show the \ECFs for cloud-free models with $T_\mathrm{eff}=$700 K (left panel) and 1300 K (right panel) and for $\log g=$5.5 (solid lines), 4.5 (dashed lines) or 3.5 (dotted lines). 
As expected, for the same $T_\mathrm{eff}$ the higher $\log g$ is, the deeper in the atmosphere we can probe. In the $J-$ band, e.g., for $T_\mathrm{eff}=$700 K we probe down to 96.5 bar for $\log g=$5.5, 25 bar for $\log g=$4.5 and 6.4 bar for $\log g=$3.5. Additionally, in models with higher $T_\mathrm{eff}$ we can probe lower pressures (i.e., higher altitudes in the atmosphere). In the $J-$ band, e.g., for $T_\mathrm{eff}=$1300 K we probe down to 49.2 bar for  $\log g=$5.5, 11.9 bar for $\log g=$4.5 and 2.4 bar for $\log g=$3.5. The cooler $T_\mathrm{eff}$ changes the atmospheric chemistry, with the 700 K spectra exhibiting $NH_3$ absorption around 10 $\mu$m, in contrast to the 1300 K model. This results in the $NH_3$ band probing lower pressures than the surrounding continuum in the 700 K model, while it probes comparable pressures to the surrounding continuum in the 1300 K model.

In Fig.~\ref{fig:LTcf_fsed2} we show the \ECFs for cloudy atmospheres with $T_\mathrm{eff}=$700 K (left panels), 1300 K (middle panels) and 1900 K (right panels). Our model atmospheres have $f_\mathrm{sed}=2$ and $\log g=$3.5 (top panels), $\log g=$4.5 (middle panels), or 5.5 (bottom panels). Here we focus on the $\log K_\mathrm{zz}=12$ case (blue lines).
The introduction of clouds in the atmosphere blocks out the larger pressures (due to increased opacity) in the 1300 K model resulting in us probing lower pressures in this model atmosphere (27.6 bar for $\log g =5.5$, 7.2 bar for $\log g =4.5$ and 1.7 bar for $\log g =3.5$ in the $J-$ band; vs 49.2 bar for $\log g =5.5$, 11.9 bar for $\log g =4.5$ and 2.4 bar for $\log g =3.5$  for the cloud-free model). In the 700 K models, we probe deeper depths  (96.5 bar for $\log g =5.5$, 25 bar for $\log g =4.5$ and 6.4 bar for $\log g =3.5$ in the $J-$ band;  which are the same depths probed for the cloud-free models) (see Sect.~\ref{Cloud Thinning} for a discussion on how high $\log K_\mathrm{zz}$ models produce cloud-free results for spectra and \ECFs).

\subsection{The effect of $K_\mathrm{zz}$}

Our model atmospheres in Fig.~\ref{fig:LTcf_fsed2} have a $\log K_\mathrm{zz}=8$ (black lines), 9 (magenta lines) and 10 (blue lines). Across the full temperature range and gravities studied here, the lower $K_\mathrm{zz}$ values result in lower pressures (higher altitudes) being probed across the full $1 -15\ \mu$m wavelength range shown. Lower $K_\mathrm{zz}$  results in more efficient condensate settling \citep{Ackerman_2001_methods}, and moves the pressures where flux originates from at lower pressures (above the cloud decks). 
For example,{ at the highest gravity  ( $\log g = 5.5$)} in the $J-$ band we probe down to 23.8, 49.3 and 96.5 bar for $\log K_\mathrm{zz}=8$, 9 and 10 respectively. In the $H-$ band we probe down to 18.4, 33.9 and 41.5 bar, and in the $K-$ band to 9.8, 11.2 and 11.6 bar.  Notably, for $\log g = 5.5$, the $\log K_\mathrm{zz} = 10$  follows the trend that high $\log K_\mathrm{zz}$ models yield results similar to cloud-free models, see discussion in Sect.~4.3, while for the $700\,\mathrm{K}$, $\log g = 3.5$ model, all $\log K_\mathrm{zz}$ values probe very similar pressures in the $K-$ band. In these models, the $K-$ band, 
as it probes lower pressures in the atmosphere, experiences reduced attenuation from clouds and molecular opacity compared to the $J-$ and $H-$ bands. As a result, the emergent flux originates from similar pressure levels across a range of atmospheric profiles. 



\subsection{The effect of $f_\mathrm{sed}$}

\begin{figure*}[htbp]
  \centering

  \includegraphics[width=0.9\textwidth]{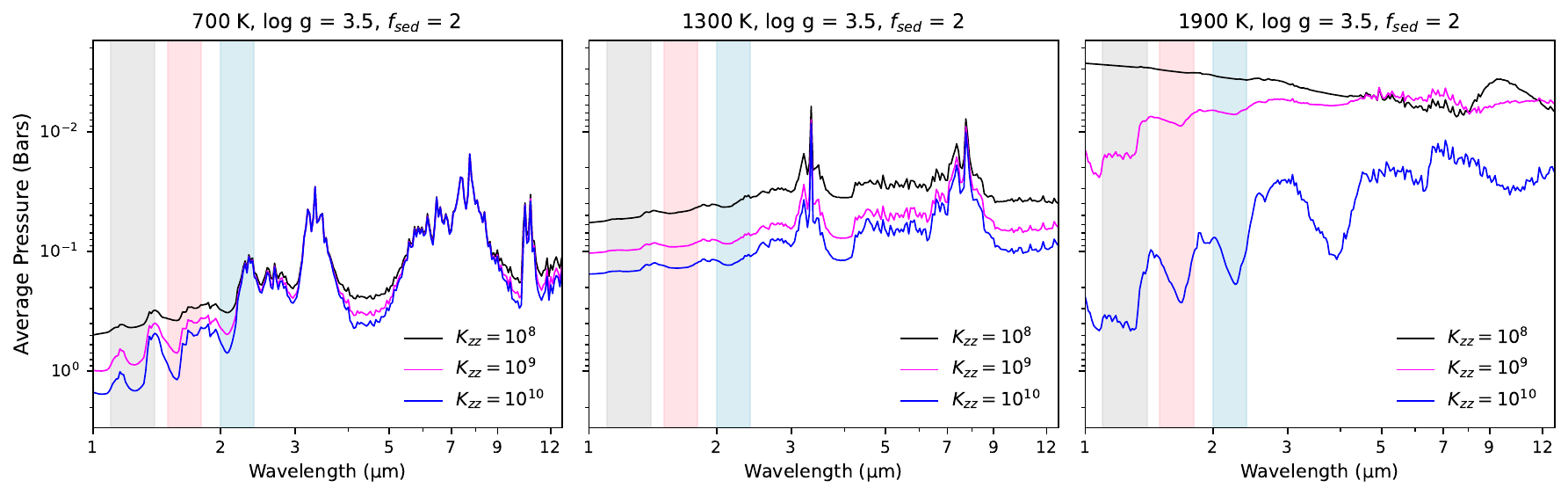}

  \vspace{-.5em} 

  \includegraphics[width=0.9\textwidth]{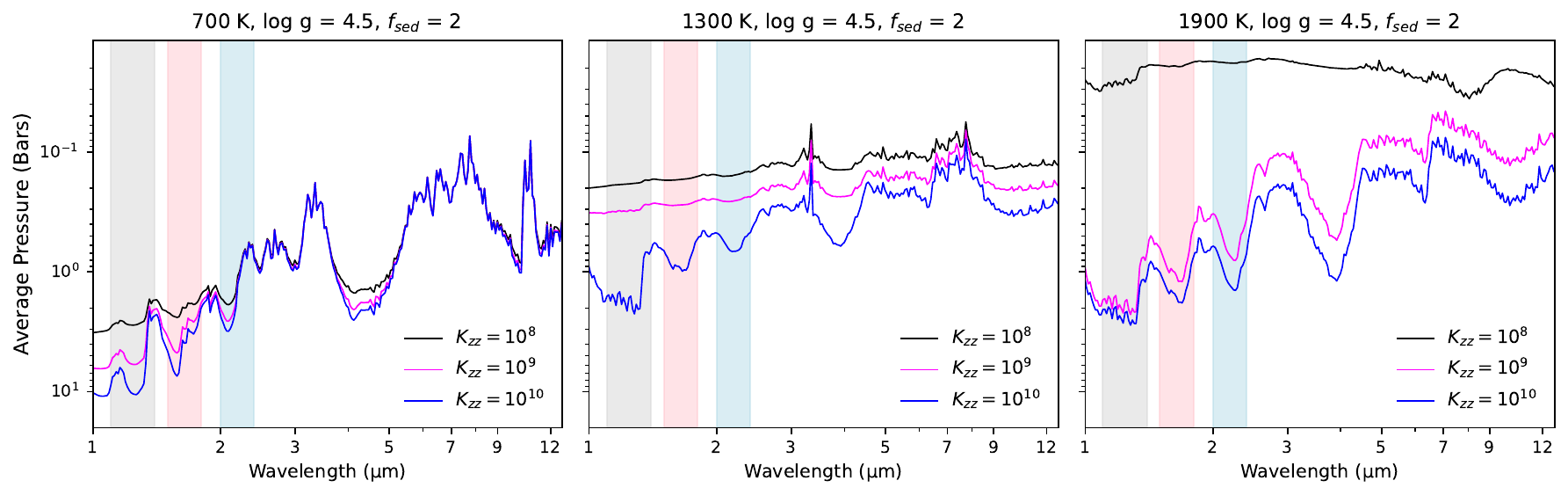}

  \vspace{-.5em} 

  \includegraphics[width=0.9\textwidth]{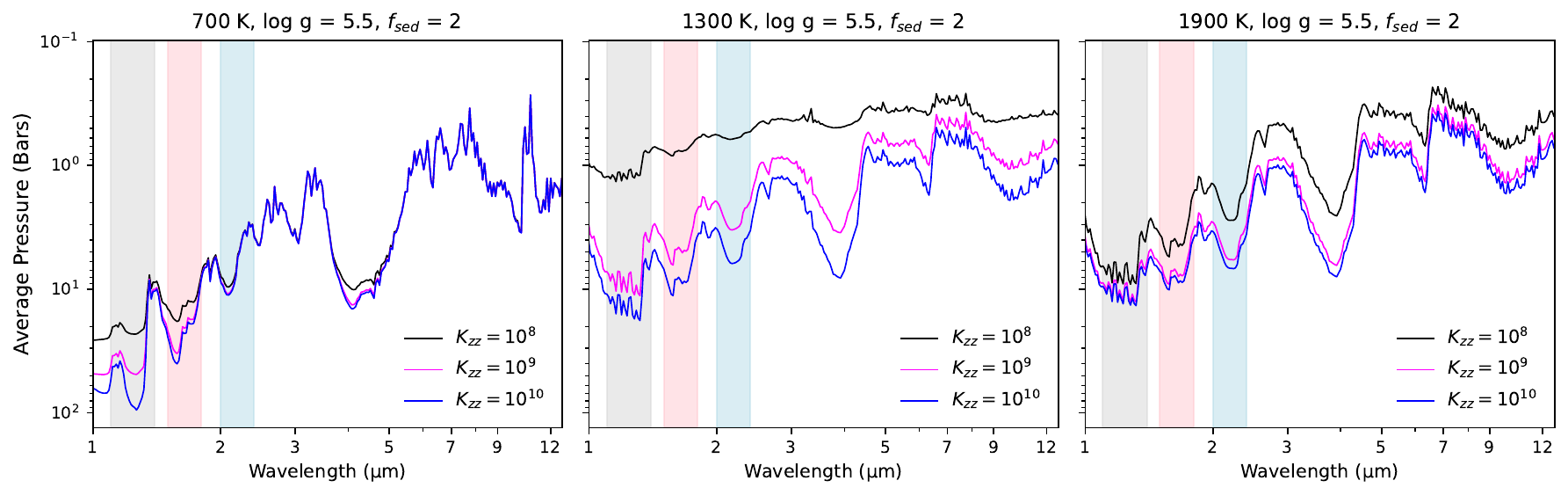}

  \caption{{Contribution functions of cloudy model atmospheres with $T_\mathrm{eff}=700$ K (left panels), 1300 K (middle panels), and 1900 K (right panels) with a gravity of $\log g =$ 3.5 (top panels), $\log g =$ 4.5 (middle panels), and $\log g =$ 5.5 (bottom panels). Our models had $f_\mathrm{sed}=2$ and $\log K_\mathrm{zz}= 8$ (black lines), 9 (magenta lines), or 10 (blue lines). As in Fig.~\ref{fig:cfree}, the $J-$, $H-$ and $K-$bands are highlighted in gray, pink and blue respectively}}
  \label{fig:LTcf_fsed2}
\end{figure*}

\begin{figure*}[htbp]
  \centering
  \includegraphics[width=0.9\textwidth]{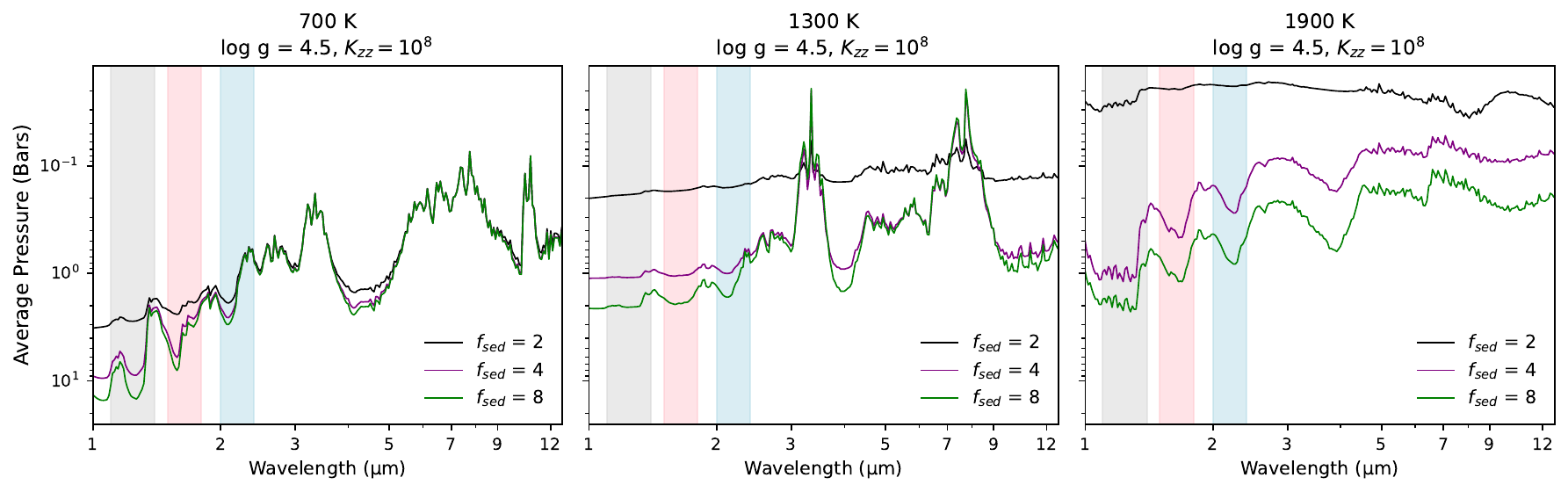}
  \caption{ 
  Contribution functions 
  for a model atmosphere with $T_\mathrm{eff}=$700 K (left panel), 1300 K (middle panel) or 1900 K (right panel), $\log g =4.5$ and $\log K_\mathrm{zz}=8$. Our atmospheres have 
$f_\mathrm{sed}=$2 (black lines), 4 (magenta lines) and 8 (blue lines). As in Fig.~\ref{fig:cfree} the shaded areas highlight the $J-$, $H-$ and $K-$ bands. As expected, we can probe deeper pressures in the colder atmosphere than the hotter atmosphere. }
  \label{fig:PressurePanel_3panel_4.5_Kzz8}
\end{figure*}

In Fig.~\ref{fig:PressurePanel_3panel_4.5_Kzz8}, our atmospheric models have $T_\mathrm{eff}=700$ K (left panel), $1300$ K (middle panel) and $1900$ K (right panel), $\log g = 4.5$, $\log K_\mathrm{zz} = 8$ and $f_\mathrm{sed}=2$ (black lines), 4 (magenta lines) and 8 (blue lines). For comparison, the black line in Fig. ~\ref{fig:PressurePanel_3panel_4.5_Kzz8} corresponds to the black line in Fig.~\ref{fig:LTcf_fsed2}. 
As expected, the colder the atmosphere is, the deeper in the atmosphere we can probe. In cloud-free regions, this is due to the smaller atmospheric scale height at lower temperatures, which compresses the atmosphere vertically. In cloudy regions, lower temperatures shift condensation to higher-pressure levels, leaving the upper atmosphere more transparent. 
When the $f_\mathrm{sed}$ is low, i.e., when the clouds are more optically thick, the atmospheric pressures probed are lower \textbf{\citep{Ackerman_2001_methods, Morley2012}}. For example, for the 1900 K model in the $J-$band we probe down to 
{0.03 bar for $f_\mathrm{sed}=2$, 1.6 bar for $f_\mathrm{sed}=4$ and 3 bar for $f_\mathrm{sed}=8$ model.} Table~\ref{tab:table_pres} gives an overview of the pressures probed in the $J-$, $H-$, and $K-$  band for our model atmospheres of Figs~\ref{fig:LTcf_fsed2} and~\ref{fig:PressurePanel_3panel_4.5_Kzz8}. Overall, 
our results show that $K_\mathrm{zz}$ variations consistently produce slightly larger changes in the depth probed in a given band than $f_\mathrm{sed}$ variations. For example, at $\log g = 4.5$, changing $\log K_\mathrm{zz}$ from 8 to 12 results in pressure changes of 88.1\% at 700 K, 97.3\% at 1300 K, and 99.1\% at 1900 K. In contrast, changing $f_\mathrm{sed}$ from 2 to 8 results in changes in the pressure probed of 80.6\% at 700 K, 90.8\% at 1300 K, and 98.9\% at 1900 K.

\subsection{Overall trends in key wavelength bands}

To quantify how cloud opacity alters the pressure probed in a few key wavelength bands, we calculated the difference between the pressure probed in the cloud-free and corresponding cloudy 
model, for models across our entire $T_\mathrm{eff}$, $\log g$ and $f_\mathrm{sed}$ space. Our cloudy models have $\log K_\mathrm{zz}=8$. Here we focus on the $J-$ (1.1-1.4 $\mu$m), $H-$ (1.5-1.8 $\mu$m) and $K-$(2.0-2.4 $\mu$m) bands, as well as the $\sim$1.4 $\mu$m and $\sim$6 $\mu$m $H_2O$ bands, the$\sim$4.2 $\mu$m and $\sim$15 $\mu$m $CO_2$ bands, the $\sim$2.4 $\mu$m and $\sim$4.7 $\mu$m $CO$ bands, the $\sim$1.7 $\mu$m and $\sim$3.3 $\mu$m $CH_4$, the $\sim$2 $\mu$m and $\sim$11 $\mu$m $NH_3$ molecular bands, and the $K$ and $Na$ doublets at the full  spectral resolution of the grid  (R$\sim$ 5000) .

\subsubsection{$J-$, $H-$ \& $K-$ bands}
\label{JKH-results}

\begin{figure*}[htbp!]
  \centering
  \includegraphics[width=0.7\textwidth]{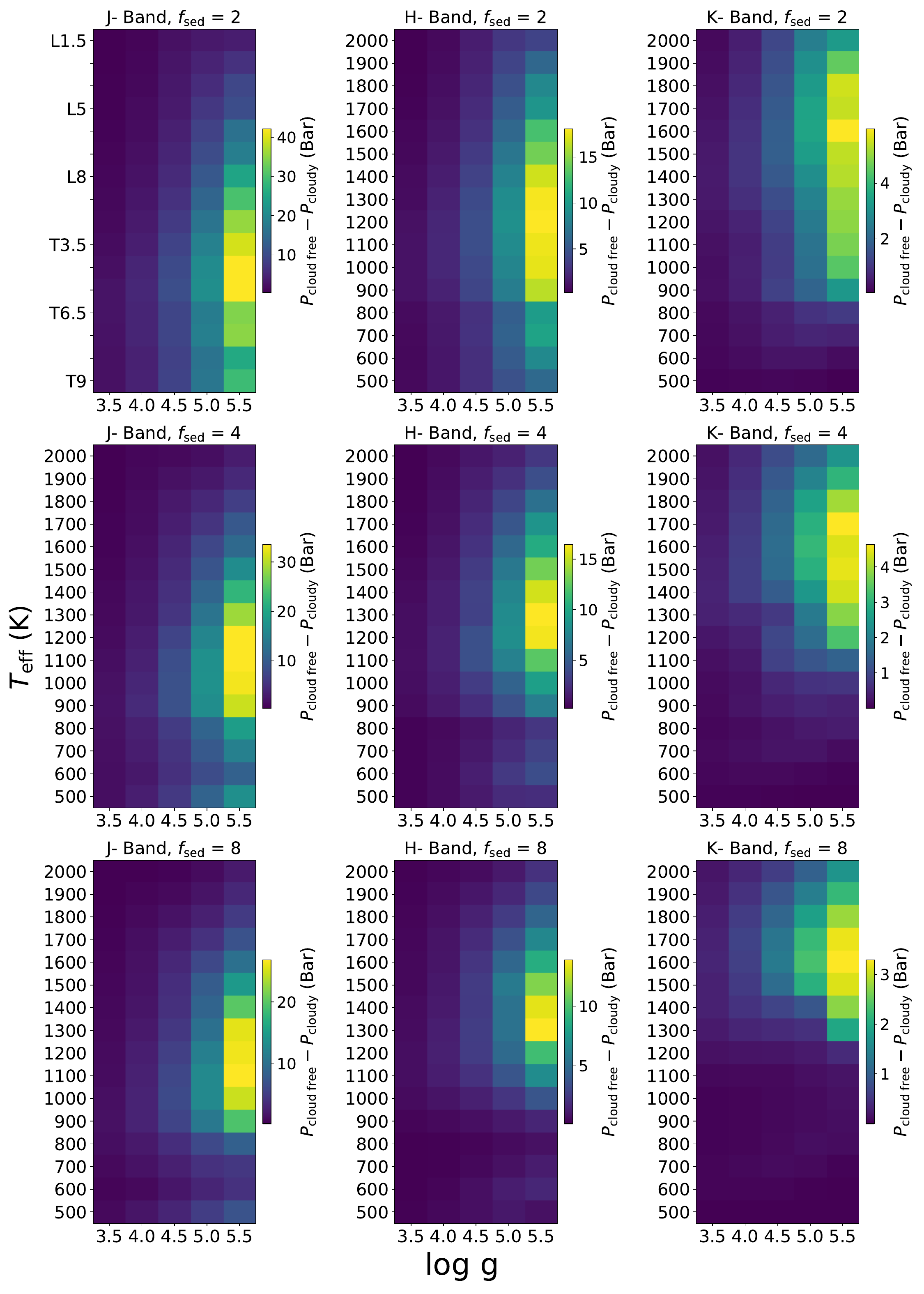}
  \caption{ The difference in the average pressure probed in the $J-$ (left panels), $H-$ (middle panels) and $K-$ (right panels) bands  between a cloudy and cloud-free atmosphere across the full range of $T_\mathrm{eff}$ and $\log g$ of our grid. The cloudy atmospheres have and $f_\mathrm{sed}$ of 2 (top row),  4 (middle row), and 8 (bottom row).
Across all $T_\mathrm{eff}$ and $\log g$ the differences are largest in the $J-$band , and the smallest in the $K-$band.  To assist comparison with observations, the first plot (top left) is labeled using the corresponding spectral type for a given temperature based on \citet{Filippazzo2015} (see also Appendix Table~\ref{tab:teff_spt_mapping}).}
  \label{fig:heatmap_results_Kzz1e8}
\end{figure*}

In Fig.~\ref{fig:heatmap_results_Kzz1e8} we plot the difference $\delta P$ and identify the maximum difference  $\delta P_\mathrm{max}$ in the pressures probed in the $J-$ (left panels), $H-$ (middle panels) and $K-$ (right panels) bands, between a cloudy and clear atmosphere across the full $T_\mathrm{eff}, \log g$ parameter space studied here. The cloudy atmosphere has an $f_\mathrm{sed}=2$ (top row), 4 (middle row) and 8 (bottom row) and a $\log K_\mathrm{zz}=8$. To assist comparison with observations, the $T_\mathrm{eff}$ axis of the top-left panel of Fig.~\ref{fig:heatmap_results_Kzz1e8} is also labeled with the corresponding spectral types from \citet{Filippazzo2015}; the mapping is given in Table~\ref{tab:teff_spt_mapping}.
For all $f_\mathrm{sed}$ the largest overall $\delta P$ occurs in the $J-$band, in which for the $f_\mathrm{sed}=2$ models, the maximum difference is $\delta P_\mathrm{max}\sim40$ bar at $T_\mathrm{eff}=900$-$1000$ K and $\log g =5.5$. As $f_\mathrm{sed}$ increases, the $\delta P_\mathrm{max}$  shifts to higher $T_\mathrm{eff}$ and decrease in magnitude, dropping to $\delta P\sim$30 bar at $f_\mathrm{sed}=4$ and  $\delta P\sim$25 bar at $f_\mathrm{sed}=8$.  

In the $H-$ and $K-$ bands, the trend is similar but less pronounced. For the $H-$ band, $\delta P$ reaches a maximum at $T_\mathrm{eff}=$1200–1300 K of $\sim18$ bar at $T_\mathrm{eff}=$1200 for the $f_\mathrm{sed}=2$ models, decreasing to $\delta P\sim$ 14 bar at $T_\mathrm{eff}=$1300 K for the $f_\mathrm{sed}=8$ models. The $K-$band shows the 
smallest differences (weakest cloud impact) 
with maximum differences $\delta P \le 6$ bar across the parameter space.

\begin{figure*}
  \centering
 \includegraphics[width=1\textwidth]{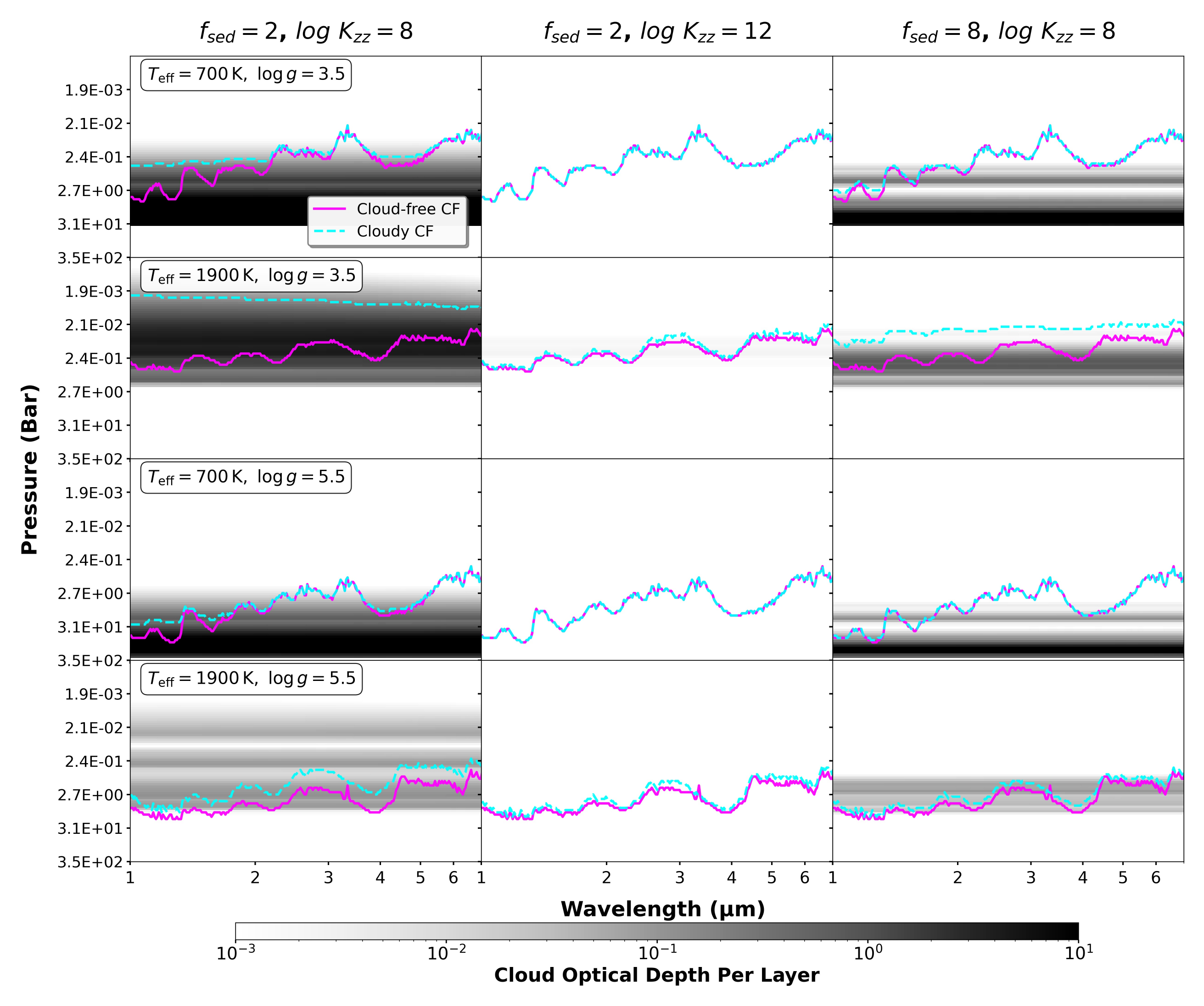}
  \caption{ Contribution functions (CF) for cloudy (cyan dashed line) and cloud-free (magenta solid line) brown dwarfs are shown for model atmospheres with $T_\mathrm{eff}=$700 K (first and third row) and 1900 K (second and fourth row), with $\log g=$ 3.5 (top two rows) and 5.5 (bottom two rows). Our models have thick clouds with $f_\mathrm{sed}=2$ (left two columns) or optically thin clouds with $f_\mathrm{sed}=8$ (third column), and $\log K_{zz}=8$ (first and third column) and 12 (second column). Also shown is the vertical extent of the clouds which, as expected, is greater for $f_\mathrm{sed}=2$ than for $f_\mathrm{sed}=8$. The second column illustrates that very strong vertical mixing ($\log K_\mathrm{zz}$=12) drives the cloudy model toward the nearly cloud-free limit $\tau < 0.035$), with contribution functions that closely resemble those of the cloud-free atmosphere.
 } 
  
  \label{fig:cldy_vs_cfree}
\end{figure*}

{For all $f_\mathrm{sed}$ and all bands, the largest $\delta P$ occurs for $\log g = 5.5$. For all $T_\mathrm{eff}$ the low gravity ($\log g =[3.5,4.0]$) models show a $\delta P \rightarrow 0$ for all bands. 
{ These trends are due to the small pressures probed in the low gravity models resulting in small absolute $\delta P$ even though the relative $\delta P$ might be upwards of 90\%, and conversely the higher pressures probed in the high gravity models.}
Additionally, $\delta P \rightarrow 0$ for the highest $T_\mathrm{eff}$ models in the $J-$band (for all $f_\mathrm{sed}$) and for the coldest models in the $K-$band. The highest $\delta P$ are noted for $T_\mathrm{eff}\geq1000$K for all bands, with the peak appearing deeper in the atmosphere in the $J-$band and higher in the atmosphere in the $H-$ and $K-$ band. }
The trend with $T_\mathrm{eff}$ is due to the fact that clouds condense at 
lower pressures with increasing $T_\mathrm{eff}$ (see, e.g., \citealp{batalha2025}) and the maximum optical thickness of the low $T_\mathrm{eff}$ models are deeper in the atmosphere than where the clear atmosphere \ECFs probe, while it is higher in the atmosphere than where the clear \ECFs probe for the high $T_\mathrm{eff}$ models  (e.g., compare the top two panels in Fig.~\ref{fig:cldy_vs_cfree}). 

{Finally, for all bands, the larger the $f_\mathrm{sed}$ is the smaller $\delta P$ is.  This is due to the reduction of the optical thickness of the clouds with increasing $f_\mathrm{sed}$, that makes the optical properties of the atmosphere resemble closer that of a cloud-free atmosphere at larger $f_\mathrm{sed}$ as shown in  Fig.~\ref{fig:cldy_vs_cfree} .
}

\begin{figure*}[htbp!]
  \centering

  \includegraphics[width=0.9\textwidth]{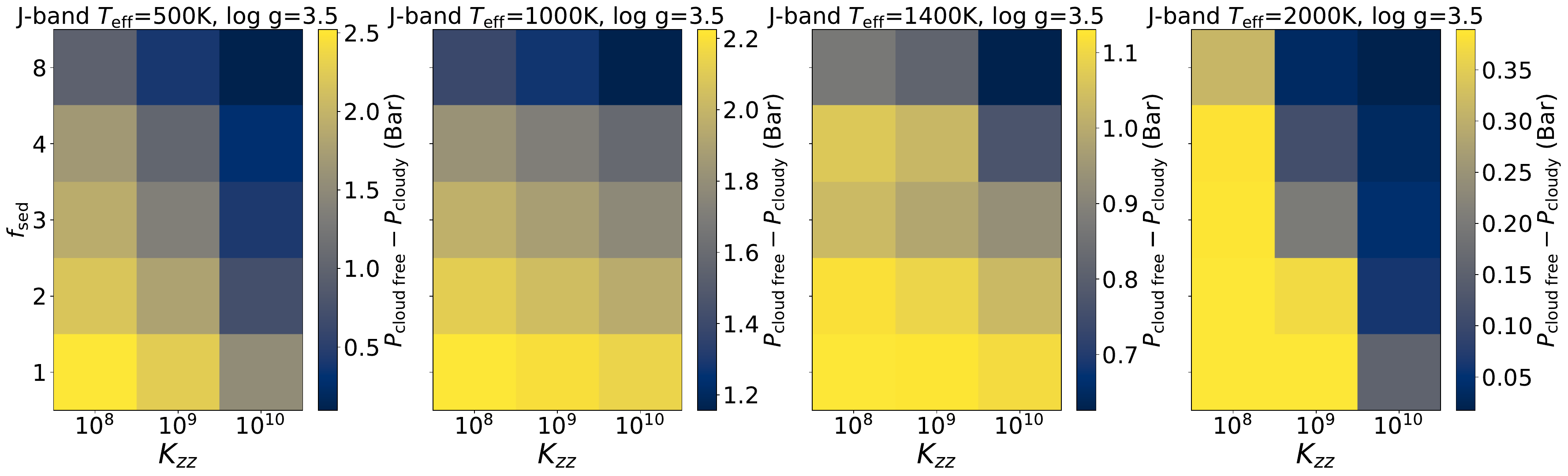}

  \vspace{-1em} 
  \hspace{1em}

  \includegraphics[width=0.9\textwidth]{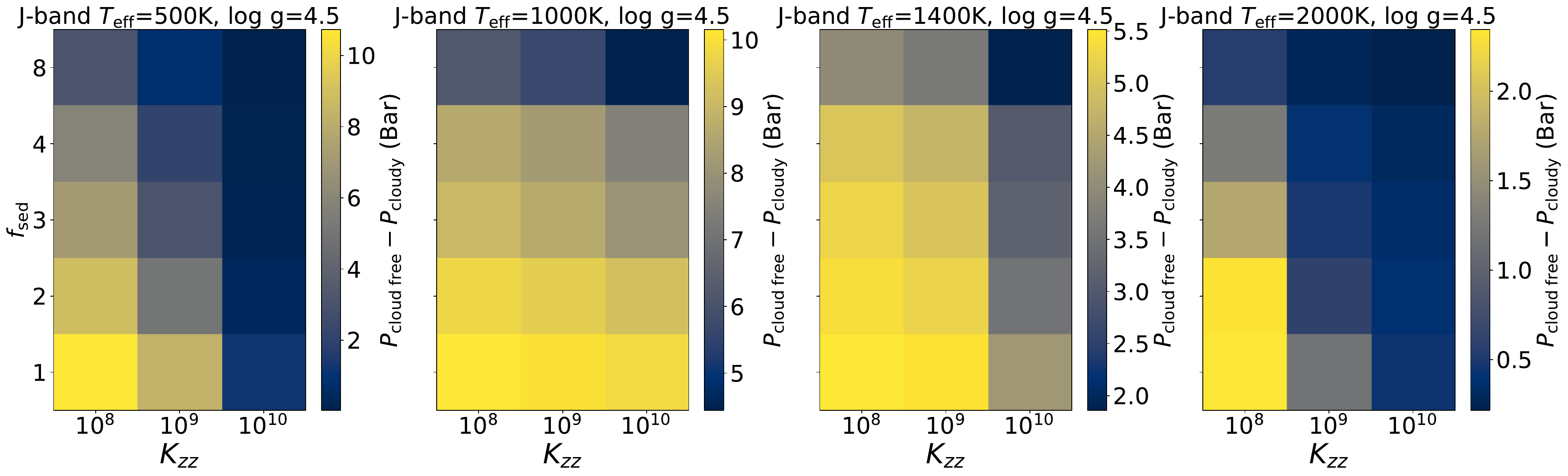}

  \vspace{-1em} 
  \hspace{1em}

  \includegraphics[width=0.9\textwidth]{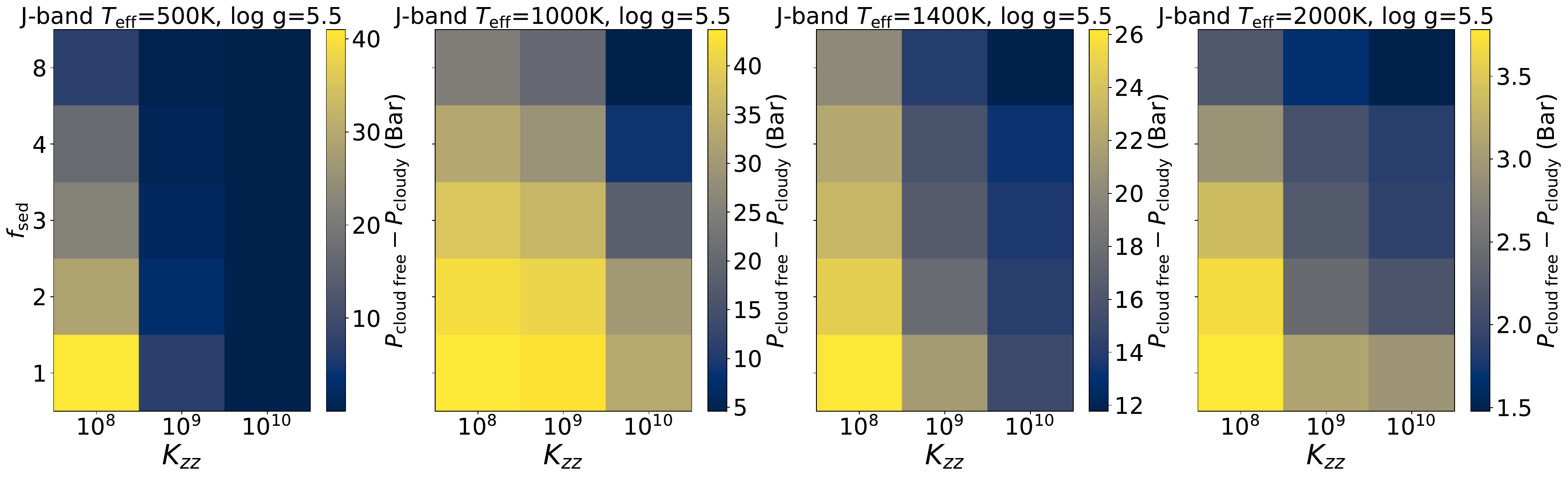}

  \caption{The difference in the average pressure probed in the $J$-band between a cloudy and cloud-free atmosphere with $T_\mathrm{eff}=500$ K (first column), 1000 K (second column), 1400 K (third column) and 2000 K (fourth column), and $\log g=$3.5 (top row), 4.5 (middle row) and 5.5 (bottom row). The model atmospheres have $f_\mathrm{sed}=1$,2,3,4 or 8 and  $\log K_\mathrm{zz}\in[8,10]$. }
  \label{fig:pressure_diff_heatmap_gravity}
\end{figure*}

{To further evaluate how cloud opacity affects the emission depth across our parameter space in the $J-$ band, which appears to be the most sensitive wavelength region to clouds, in Fig.~\ref{fig:pressure_diff_heatmap_gravity} we plot the difference $\delta P$ between the pressures probed by cloudy and the corresponding cloud-free profiles as a function of $K_\mathrm{zz}$ and $f_\mathrm{sed}$. Here we focus on $T_\mathrm{eff}=$500 K (first column), 1000 K (second column), 1400 K (third column), and 2000 K (fourth column) and $\log g=$3.5 (top row), 4.5 (middle row), and 5.5 (bottom row). As expected, for all gravities the strongest differences occur for the lower $f_\mathrm{sed}$ (thicker clouds). Thick cloud models  with moderate-to-low $K_\mathrm{zz}$  ($f_\mathrm{sed}=$1 and a $\log K_\mathrm{zz}=8$) appears to show the  largest $\delta P$ for all $T_\mathrm{eff}, \log g$.} 
{Peak $\delta P\gtrsim$40 bars at 1000 K and $\log g=$5.5, while remains below 2.5 bars at $\log g=$ 3.5 for the same temperature (see Fig.~\ref{fig:pressure_diff_heatmap_gravity}). At 1400 K and $\log g=$5.5, $\delta P_\mathrm{max}\sim26$ bars but declines at high-$K_\mathrm{zz}$, high-$f_\mathrm{sed}$ values ($\log K_\mathrm{zz}=11$ or $12$).  
For $T_\mathrm{eff}=$2000 K across all gravities $\delta P$ is low, and $\delta P\rightarrow 0$ for $\log g=$3.5. 
}

\subsubsection{$H_2O$ Bands }

\begin{figure}[htbp!]
  \centering
  \includegraphics[width=\linewidth]{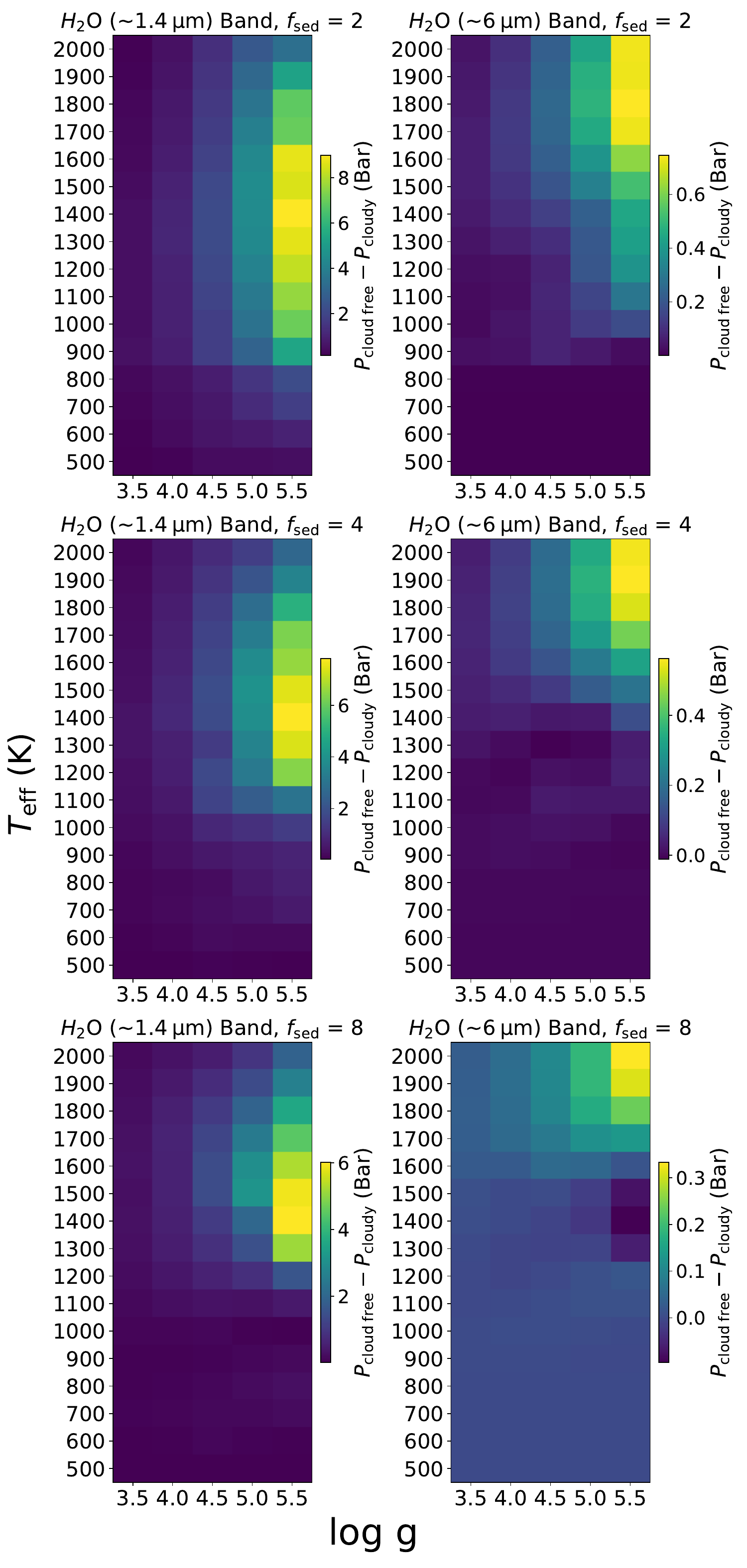}
  \caption{Similar to Fig.~\ref{fig:heatmap_results_Kzz1e8} but for the $H_2O$ bands around 1.4 $\mu$m (right panels) and 6 $\mu$m (left panels).}
  \label{fig:heatmap_results_H2O_Kzz1e8}
\end{figure}

H$_2$O is one of the dominant molecules in L and T dwarfs, with broad absorption bands that intensify as objects cool \citep[e.g.,][]{Marley_2021,
McLean2003, Geballe2001}. Fig.~\ref{fig:heatmap_results_H2O_Kzz1e8} shows $\delta P$ for our models in two $H_2O$ bands: at 1.4 $\mu$m (left panels) and 6 $\mu$m (right panels). The 1.4 $\mu$m band shows greater sensitivity to the cloud opacity than the 6$\mu$m band. For the  $f_\mathrm{sed}=2$ models in the 1.4$\mu$m band $\delta P$ peaks at $\sim9$ bars for $T_\mathrm{eff}=1400$K and $\log g =5.5$. As $f_\mathrm{sed}$ increases to 4 and 8, $\delta P$ decreases to $\sim7$ and $\sim6$ bars respectively. In contrast, the 6$\mu$m band shows a weaker cloud sensitivity, with $\delta P \le$ 0.6 bars, even in the thickest cloud cases. For $f_\mathrm{sed}=8$, cloud effects in the 6$\mu$m band are nearly negligible across the entire $T_\mathrm{eff}-\log g$ grid.  
{Similar to the $J-$, $H-$, $K-$ band trends, the difference in the sensitivity of the 1.4 $\mu$m and 6 $\mu$m bands is due to the different pressures (deeper vs higher in the atmosphere) that the two bands probe. The pressures probed in the 1.4 $\mu$m band are overall deeper or around the pressure where the larger cloud opacity is seen (see, e.g., Fig.~\ref{fig:cldy_vs_cfree}). On the other hand, the pressures probed in the 6 $\mu$m band are around the upper parts, or even above the cloud layer (see, e.g., Fig.~\ref{fig:cldy_vs_cfree}) 
}

{
Taking into account the variability of $\delta P$ with $K_\mathrm{zz}$ and $f_\mathrm{sed}$ (see first row in Fig.~\ref{fig:pressure_diff_heatmap_molecules}),$\delta P_\mathrm{max, H_2O}$ reaches 1 to 2 bars for $T_\mathrm{eff}\in[1000,2000]$ K. }
{This shows that the $H_2O$ bands remain consistently sensitive to cloud opacity over a wide temperature range.
}

\subsubsection{$CO_2$ Bands}

\begin{figure}[htbp!]
  \centering
  \includegraphics[width=\linewidth]{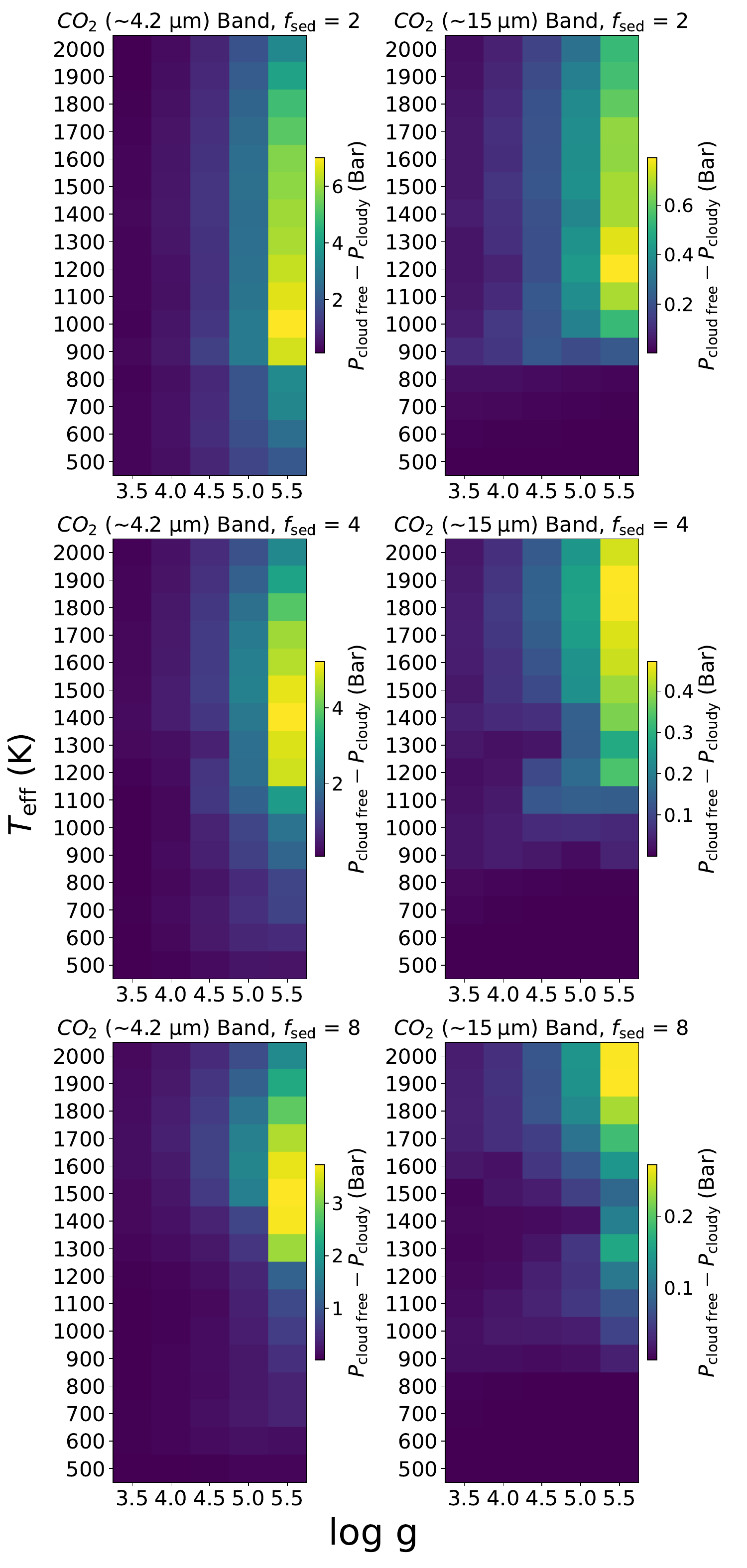}
  \caption{Similar to Fig.~\ref{fig:heatmap_results_Kzz1e8} but for the $CO_2$ bands around 4.2 $\mu$m (right panels) and 15 $\mu$m (left panels).}
  \label{fig:heatmap_results_CO2_Kzz1e8}
\end{figure}

$CO_2$ absorption is negligible in L dwarfs and early T dwarfs, but it appears in late-T dwarfs. Although $CO_2$ is not a major opacity source compared to $H_2O $ or $CH_4$, its detection serves as a sensitive probe of atmospheric C/O chemistry and indicates extremely low temperatures where $CO_2$ can persist in equilibrium amounts \citep{Yamamura2010, Beiler2024}
In Fig.~\ref{fig:heatmap_results_CO2_Kzz1e8} we show $\delta P$ for the $CO_2$ bands at 4.2 $\mu$m (left panels) and 15 $\mu$m (right panels). In the 4.2 $\mu$m band, the largest pressure difference ($\delta P\sim6$ bars) appears around $T_\mathrm{eff}=$1000 K and $\log g=5.5$ for the  $f_\mathrm{sed}=2$ models. For $f_\mathrm{sed} =$ 4 and 8
$\delta P_\mathrm{max}\sim6$ bars and $\sim4$ bars, respectively. The 15 $\mu$m $CO_2$ band exhibits a similar pattern, but with lower sensitivity: $\delta P$ remains below $\sim8$ bars across the entire $T_\mathrm{eff}-\log g$ grid. 

{
Considering $K_\mathrm{zz}$ and $f_\mathrm{sed}$ trends (see third row in Fig.~\ref{fig:pressure_diff_heatmap_molecules}),  the $CO_2$-band has a $\delta P_\mathrm{max}\sim1.5$ bar for $T_\mathrm{eff}\in[1000,1400]$ K, with a strong dependence on $f_\mathrm{sed}$ and a weaker dependence on $K_\mathrm{zz}$. As expected, the effect diminishes by 2000 K in contrast to $CO$.}

\subsubsection{$CO$ Bands }\label{subsec:cobands}

\begin{figure}[htbp!]
  \centering
  \includegraphics[width=\linewidth]{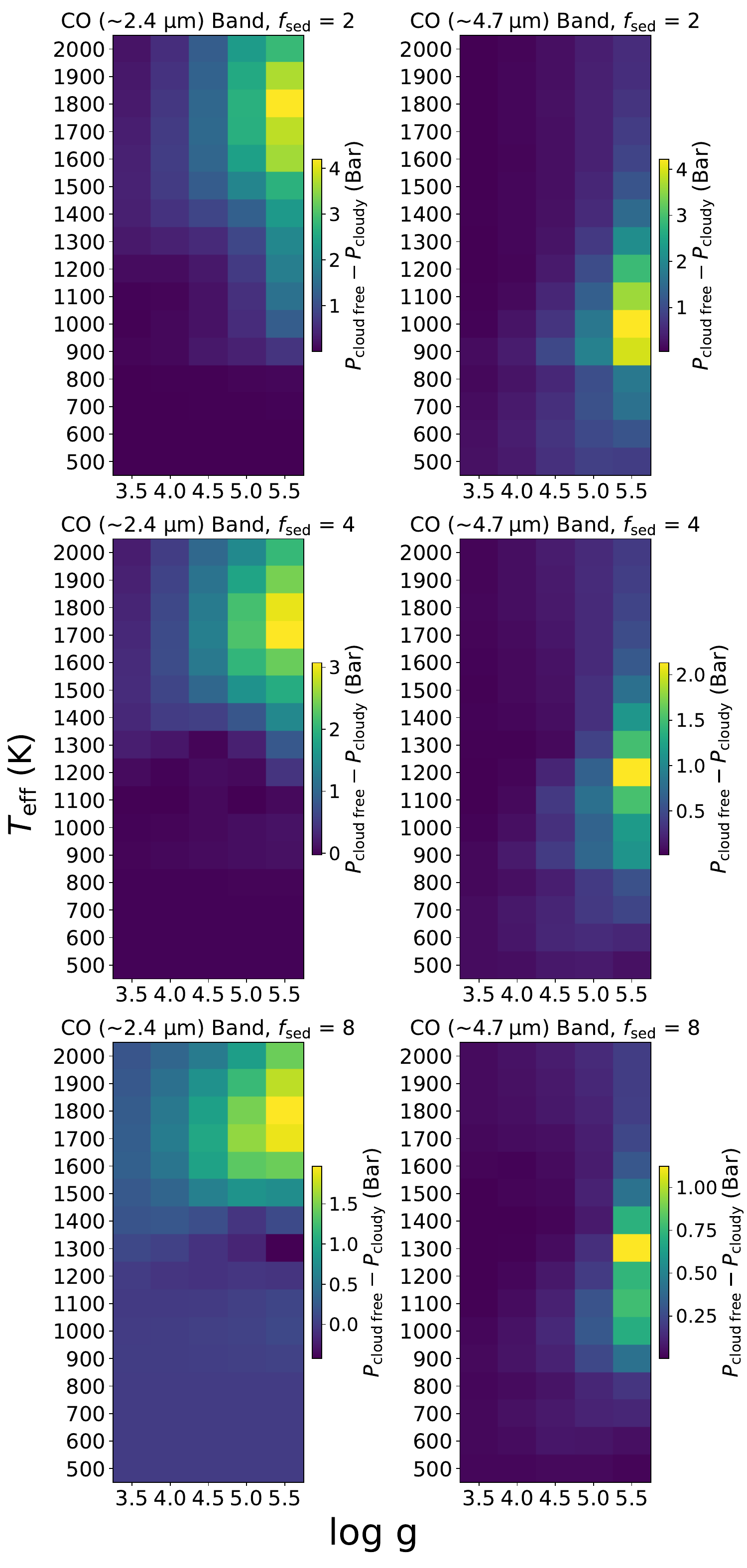}
  \caption{Similar to Fig.~\ref{fig:heatmap_results_Kzz1e8} but for the $CO$ bands and 2.4 $\mu$m (left panels) and 4.7 $\mu$m (right panels).}
  \label{fig:heatmap_results_CO_Kzz1e8}
\end{figure}

L dwarfs show strong $CO$ absorption 
but as $T_\mathrm{eff}$ decreases through the L/T transition, $CO$ features weaken while $CH_4$ bands strengthen \citep{Geballe2001}. By the mid-T dwarfs, $CO$ would be mostly converted to $CH_4$ in chemical equilibrium; however, $CO$ has been observed in T dwarf spectra 
indicating significant disequilibrium chemistry in these atmospheres \citep{Noll1997, Burrows1999}.  However, these models do not account for disequilibrium chemistry.  In Fig.~\ref{fig:heatmap_results_CO_Kzz1e8} we show $\delta P$ for the $CO$ absorption bands at 2.4 $\mu$m (left panels) and 4.7 $\mu$m (right panels). Similar to the trends we saw for the other bands, the 2.4 $\mu$m band shows stronger cloud sensitivity than the 4.7 $\mu$m band, with a peak $\delta P\sim4$bar occurring for the  $T_\mathrm{eff}=$1800 K, $\log g=$5.5,$f_\mathrm{sed}=$2 model. As $f_\mathrm{sed}$ increases to 4 and 8, $\delta P_\mathrm{max}$ declines to $\sim3$ bars and $\sim2$ bars,. For the 4.7 $\mu$m band $\delta P_\mathrm{max}\sim4$bar for the  $T_\mathrm{eff}=$1000 K, $\log g=$5.5, $f_\mathrm{sed}=$2 model.  decreasing to $\sim1$ bar at an $f_\mathrm{sed}$ of 8 ($T_\mathrm{eff}=$1200 K and 1300 K respectively).

{Considering $K_\mathrm{zz}$ and $f_\mathrm{sed}$ trends  (see fourth row in Fig.~\ref{fig:pressure_diff_heatmap_molecules}), the $CO-$band also shows a modest cloud sensitivity, with $\delta P_\mathrm{max}$ $\sim1$ bar for $T_\mathrm{eff}\in[1400,2000]$ K. $\delta P$ is larger at higher temperatures, especially for $f_\mathrm{sed}\in[1,2]$ and $\log K_\mathrm{zz}=10$.  This is due to the stronger role that $CO$ plays in hotter atmospheres where it becomes more abundant.
}

\subsubsection{$CH_{4}$ Bands}
\begin{figure}[htbp!]
  \centering
  \includegraphics[width=\linewidth]{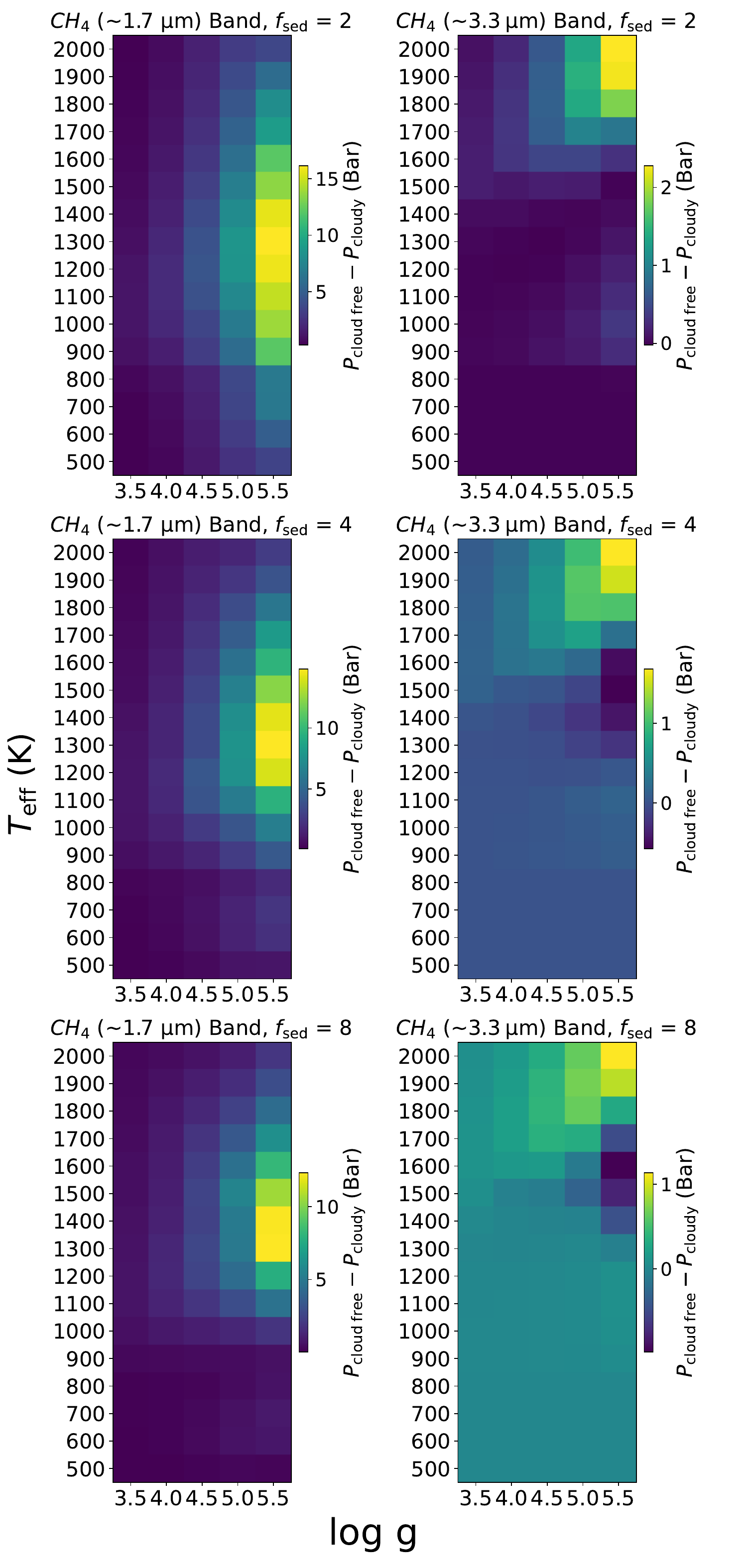}
  \caption{Similar to Fig.~\ref{fig:heatmap_results_Kzz1e8} but for the $CH_4$ bands around 1.7 $\mu$m (left panels) and 3.3 $\mu$m (right panels). }
  \label{fig:heatmap_results_CH4_Kzz1e8}
\end{figure}

As discussed in Sect.~\ref{subsec:cobands}, the onset and growth of  $CH_4$ features serves as a spectroscopic indicator of lower temperatures in  a brown dwarf atmosphere. In Fig.~\ref{fig:heatmap_results_CH4_Kzz1e8} we show $\delta P$ for $CH_4$ absorption bands at 1.7 $\mu$m (left panels)and 3.3 $\mu$m (right panels). As with the previous bands, the 1.7 $\mu$m band probes deeper layers in the atmosphere and thus shows a larger sensitivity to cloud opacity, with $\delta P_\mathrm{max}\sim15$ bar at $T_\mathrm{eff}\sim$1200–1400 K and $\log g=$5.5 for the  $f_\mathrm{sed}=2$ model. $\delta P_\mathrm{max}$ drops to $\sim15$ bars at $f_\mathrm{sed}=4$ and  $\sim12$ bar at $f_\mathrm{sed}=$8. Similarly, the 3.3 $\mu$m band has a peak $\delta P_\mathrm{max}\sim2$ bar for the $f_\mathrm{sed}=2$ model and falls to $\lesssim1$ bar for the $f_\mathrm{sed}=8$ model. Note that for the 3.3 $\mu$ m $CH_4$ band we also encounter for the first time negative $\delta P$, which indicates that the cloudy profile probes deeper pressures than the clear atmosphere profile for some of the $f_\mathrm{sed}=4$ and 8 models.  

\begin{figure*}[htbp!]
  \centering

  \includegraphics[width=0.9\textwidth]{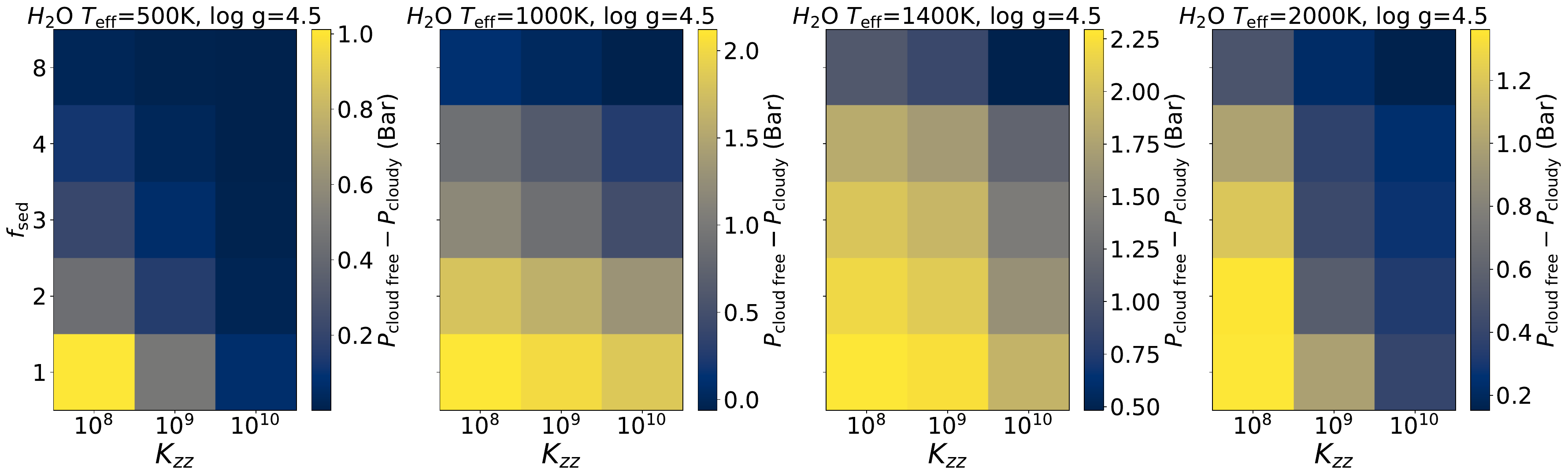}

  \vspace{-1em} 
  \hspace{1em}

  \includegraphics[width=0.9\textwidth]{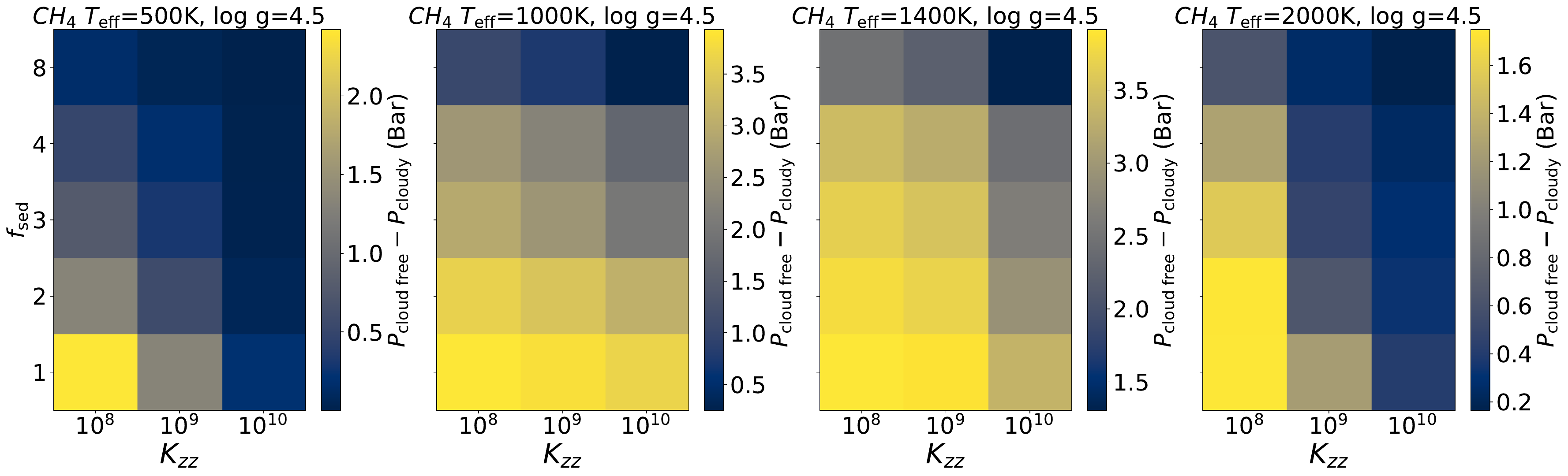}

  \vspace{-1em} 
  \hspace{1em}

  \includegraphics[width=0.9\textwidth]{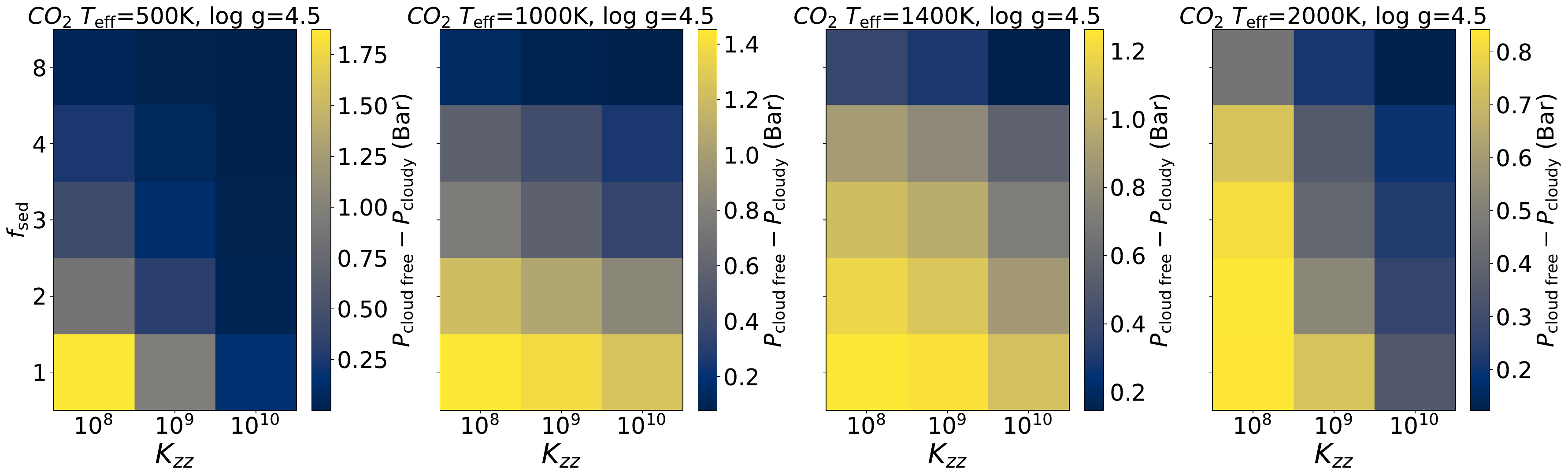}

  \vspace{-1em} 
  \hspace{1em}

  \includegraphics[width=0.9\textwidth]{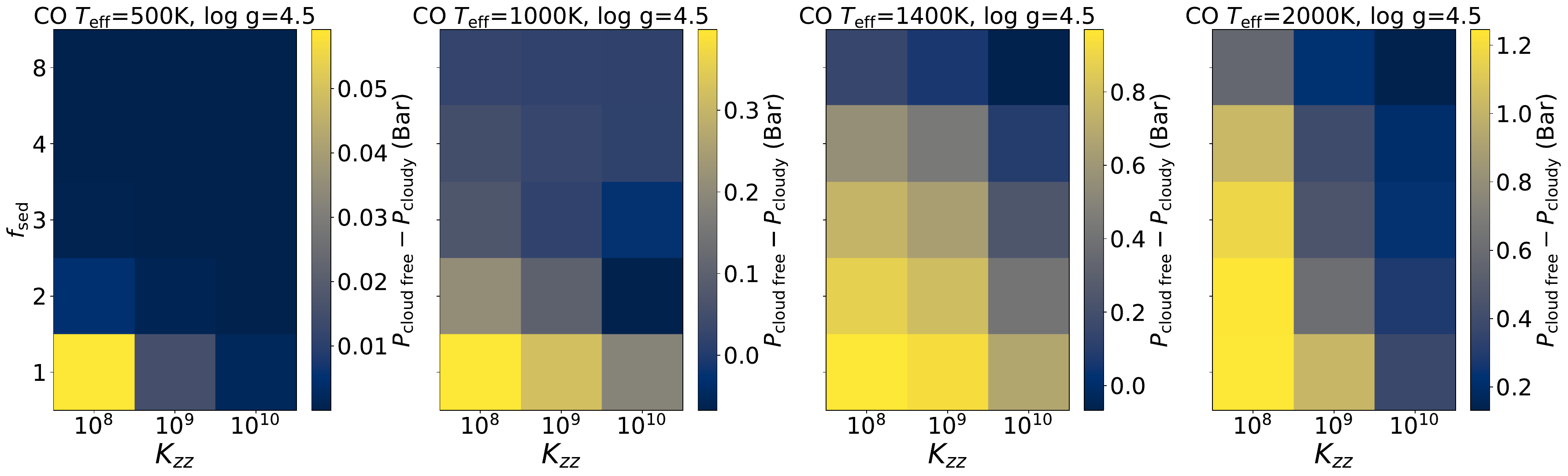}
  \hspace{1em}

  \caption{Similar to Fig.~\ref{fig:pressure_diff_heatmap_gravity} but for $H_2O$ (first row), $CH_4$ (second row) and $CO_2$ (third row) and $CO$ (fourth row).}
  \label{fig:pressure_diff_heatmap_molecules}
\end{figure*}

{Considering the $K_\mathrm{zz}$ and $f_\mathrm{sed}$ trends,  the $CH_4$-band shows strong cloud sensitivity, particularly at $T_\mathrm{eff}\in[1000,1400]$ K. $\delta P_\mathrm{max}=4$ bars for thick clouds and the $\log K_\mathrm{zz}=$ 9 and 10 models (see second row in Fig.~\ref{fig:pressure_diff_heatmap_molecules}). At $T_\mathrm{eff}=500$ K, $CH_4$  differences are smaller and $\delta P\rightarrow0$ at $T_\mathrm{eff}=2000$ K. This reflects the role of $CH_4$  in mid-to-late T dwarfs, where clouds significantly alter the observed flux.}


\subsubsection{$NH_{3}$ Bands } \label{subsubsec:NH3}
\begin{figure}[htbp!]
  \centering
  \includegraphics[width=\linewidth]{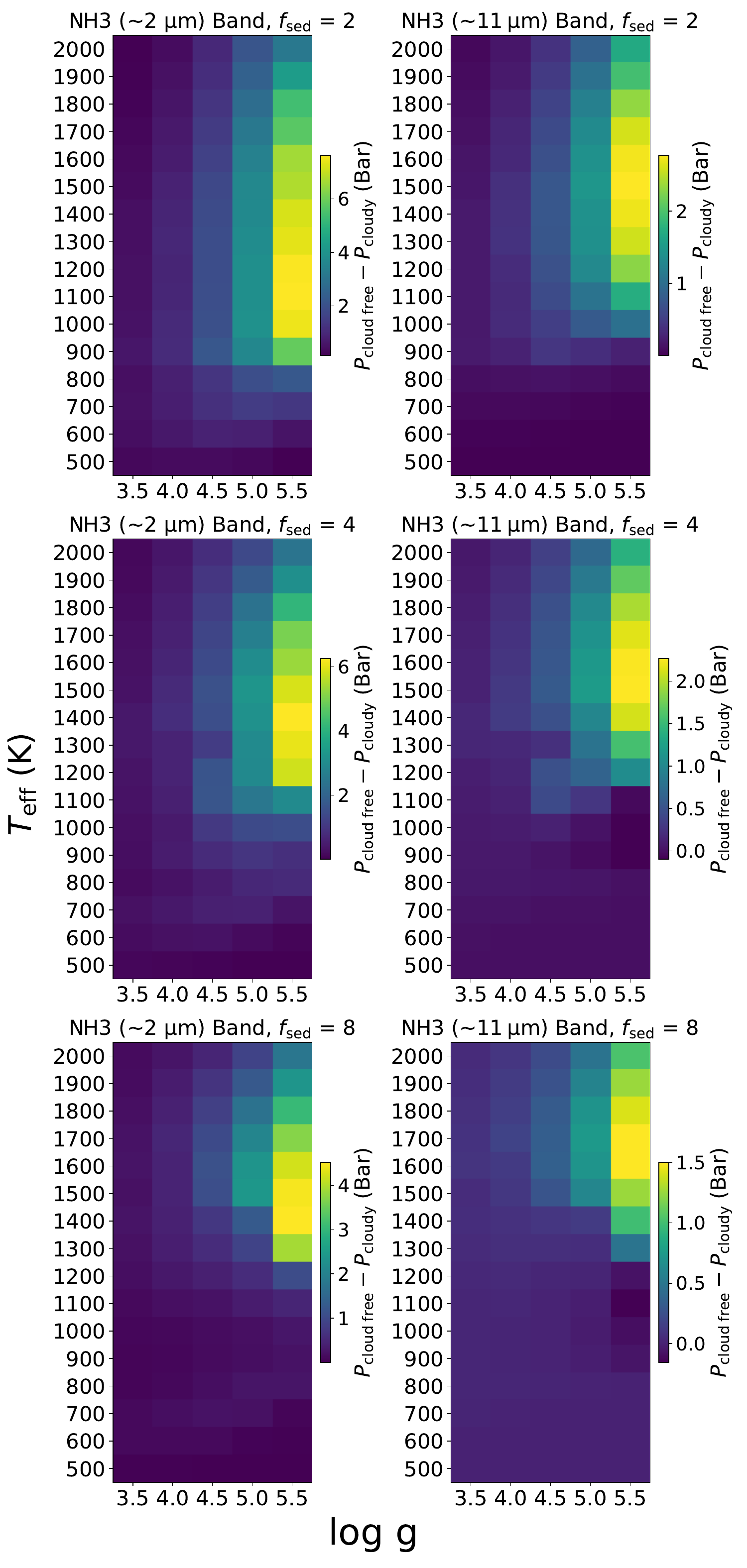}
  
 \caption{Similar to Fig.~\ref{fig:heatmap_results_Kzz1e8} but for the $NH_3$ bands around 2 $\mu$m (left panels) and 11 $\mu$m (right panels).}
  \label{fig:heatmap_results_NH3_fsed2}
\end{figure}

NH$_3$ is a sensitive indicator of brown dwarf temperature. At higher $T_{\rm eff}$ nitrogen resides mainly in N$_2$, which is spectroscopically inactive in the NIR/MIR \citep[see, e.g.,][for an in-depth discussion]{Lodders2002, Saumon2012}, while at lower  
$T_{\rm eff} (\sim 800$ K) the equilibrium shifts towards NH$_3$ \citep{Saumon2012}. Thus, while $NH_3$ absorption is negligible in L dwarfs it appears in late T dwarf spectra \citep[e.g.][]{Lodders2002,Saumon2012,Cushing2006,Beale2017} serving as an indicator of the brown dwarf temperature.
Fig.~\ref{fig:heatmap_results_NH3_fsed2} shows $\delta P$ for the $NH_3$ absorption bands around $\sim2\mu$m (left panels) and $\sim 11\mu$m (right panels). Overall, $NH_3$ shows weaker sensitivity to the clouds. The $\sim2\mu$m band shows a $\delta P_\mathrm{max}\sim6$-7 bar for the $T_\mathrm{eff}=1100$-1200 K models, with $\log g$ = 5.5 and $f_\mathrm{sed}=2$. As $f_\mathrm{sed}$ increases to 4 and 8  $\delta P_\mathrm{max}$ shifts to $T_\mathrm{eff}=1400$-1500 K. The 11 $\mu$m $NH_3$ band probes lower pressures in the atmosphere, probing the upper layers of the clouds or even above the cloud for the lower gravities, and $\delta P$ is thus less sensitive to clouds in this band than the $\sim2\mu$m band, with $\delta P\lesssim3$ bar for all $f_\mathrm{sed}$.


{Finally, considering the $K_\mathrm{zz}$ and $f_\mathrm{sed}$ trends (see Sect.~\ref{subsubsec:NaK}), the $NH_3-$ bands show low but consistent sensitivity to clouds across the grid. $\delta P_\mathrm{max}$ remains under 0.5 bars even at 500 K, with a slight increase at intermediate temperatures ($T_\mathrm{eff}\in[1000, 1400]$ K). 
}

\subsubsection{Na and K Doublets } \label{subsubsec:NaK}

\begin{figure}[htbp!]
  \centering
  \includegraphics[width=\linewidth]{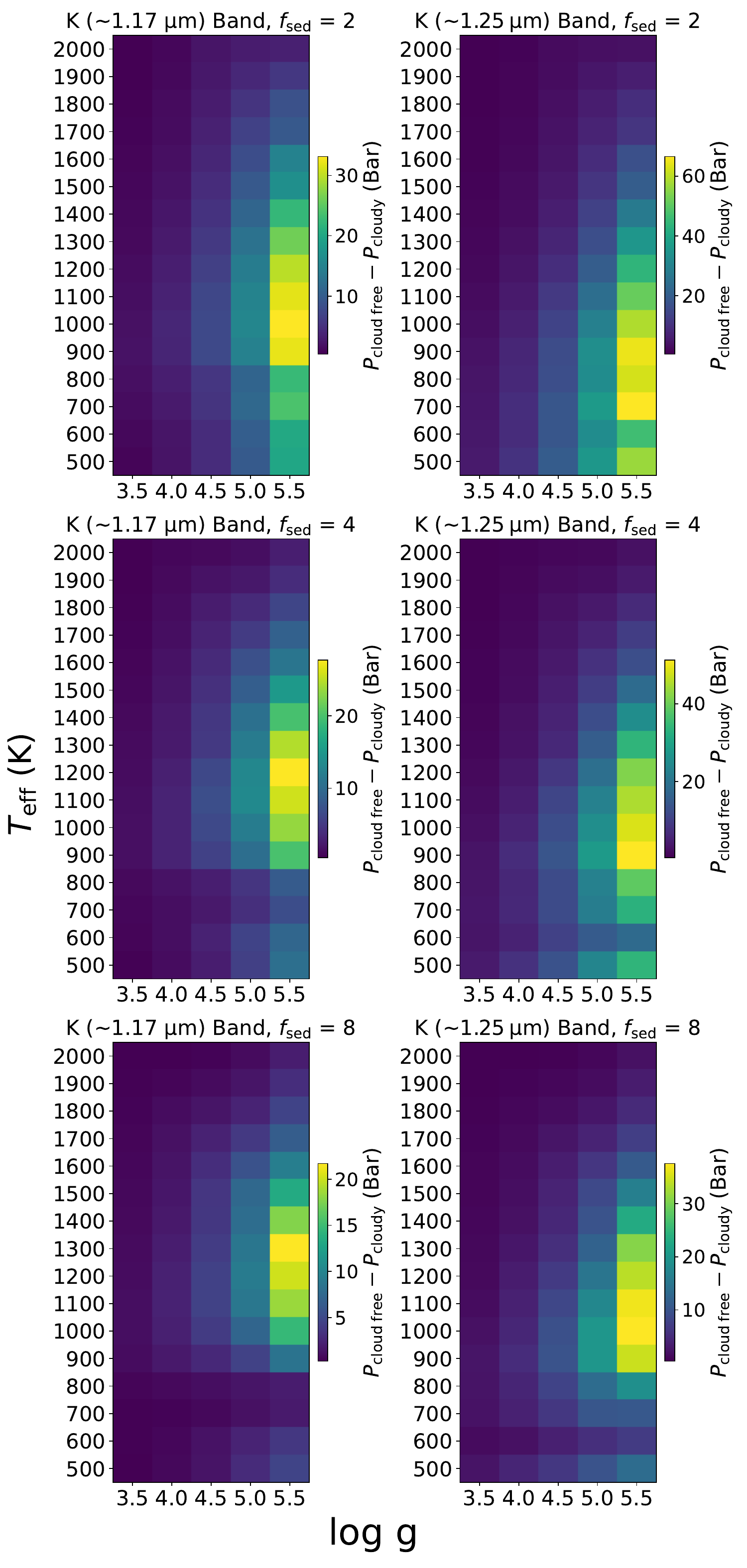}
  \caption{Similar to Fig.~\ref{fig:heatmap_results_Kzz1e8} but for the K doublets around 1.17 $\mu$m (left panels) and 1.25 $\mu$m (right panels).}
  \label{fig:heatmap_results_K_fsed2}
\end{figure}

\begin{figure}[htbp!]
  \centering
  \includegraphics[width=.6\linewidth]{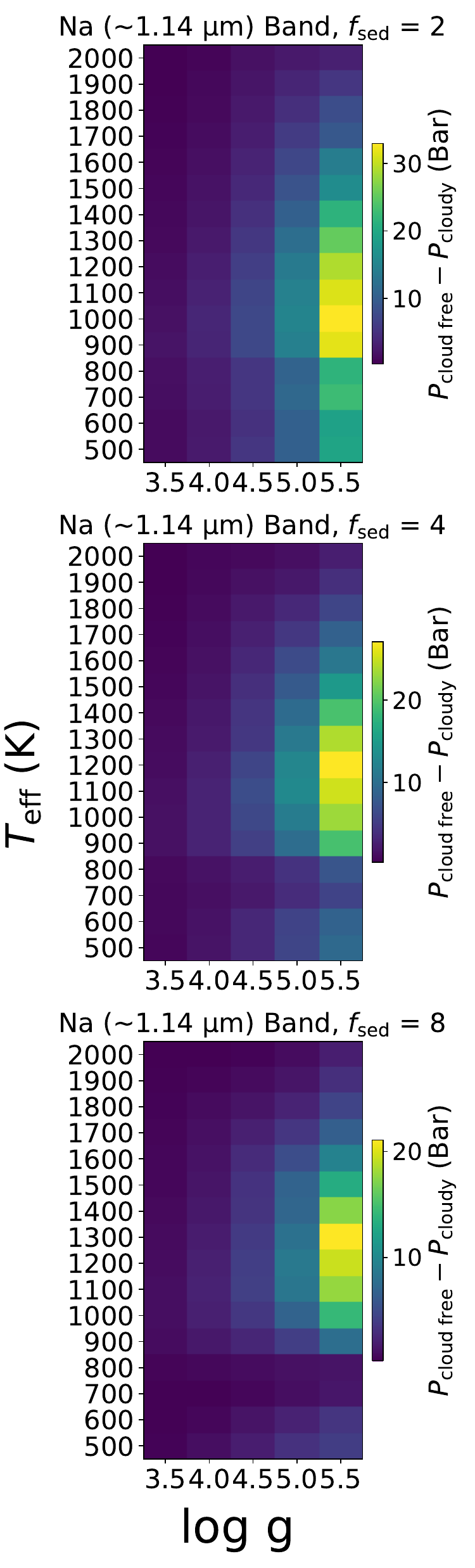}
  \caption{Similar to Fig.~\ref{fig:heatmap_results_Kzz1e8} but for the Na doublet around 1.14 $\mu$m.}
  \label{fig:heatmap_results_Na_fsed2}
\end{figure}

\begin{figure*}[htbp!]
  \centering

  \includegraphics[width=0.9\textwidth]{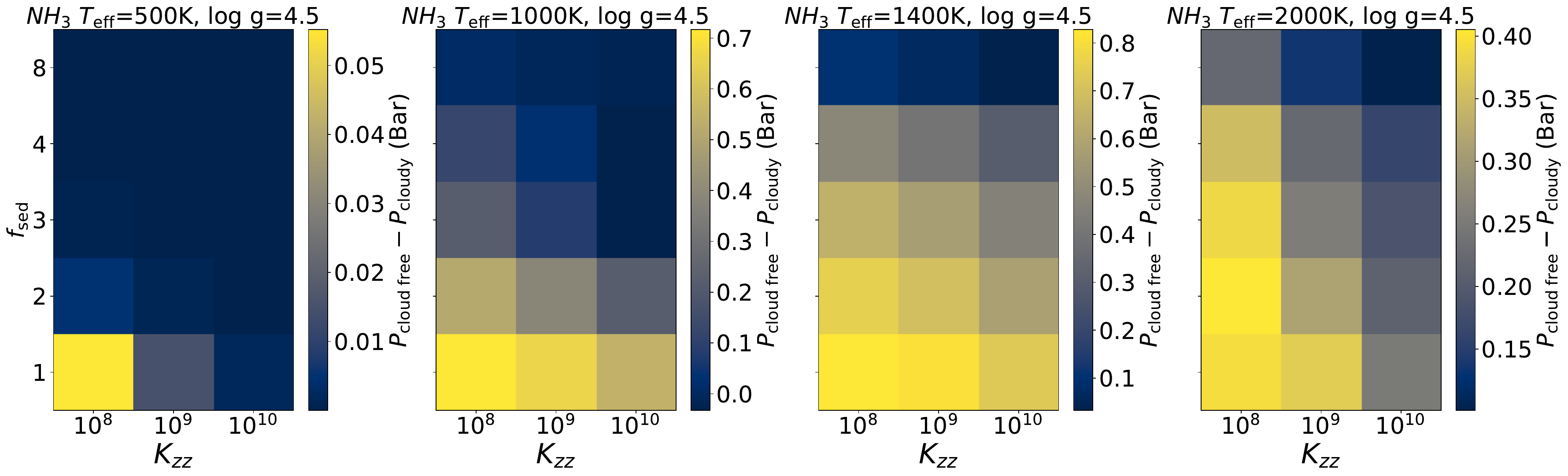}

  \vspace{-1em} 
  \hspace{1em}

  \includegraphics[width=0.9\textwidth]{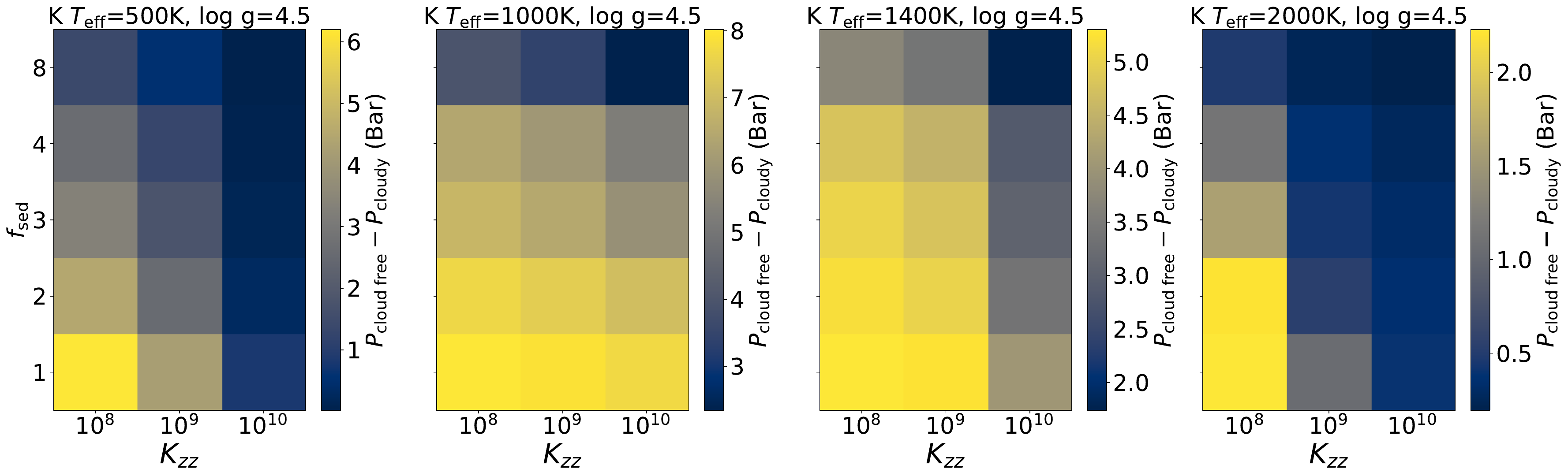}

  \vspace{-1em} 
  \hspace{1em}

  \includegraphics[width=0.9\textwidth]{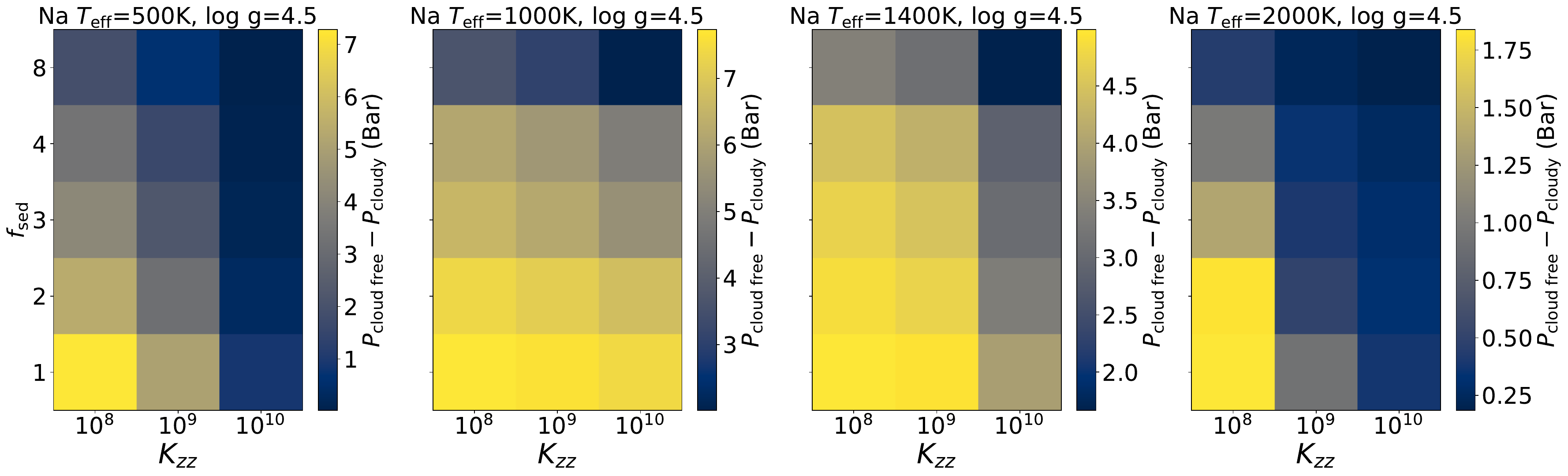}

  \caption{Similar to Fig.~\ref{fig:pressure_diff_heatmap_gravity} but for $NH_3$ (top row), $K$ (middle row) and $Na$ (bottom row). }
  \label{fig:wide}
\end{figure*}

Broadened $Na$ I and $K$ I absorption persists into the early and mid T dwarfs, and serves as a powerful diagnostic of atmospheric pressure and surface gravity in brown dwarfs ranging from $\sim$800 K to $\sim$2000 K \citep{Basri2000,Burrows2003}. $Na$ and $K$ exhibit some of the strongest cloud sensitivity in the grid, particularly in the NIR (near-infrared). In Figs.~\ref{fig:heatmap_results_K_fsed2} and ~\ref{fig:heatmap_results_Na_fsed2} we shows $\delta P$ for the 1.17 $\mu$m and the 1.25 $\mu$m $K$ doublets and the 1.14 $\mu$m $Na$ doublet. The 1.25 $\mu$m $K$ doublet shows a $\delta P_\mathrm{max}\gtrsim60$ bar for the $T_\mathrm{eff}=700$-900 K, $\log g=5.5$, $f_\mathrm{sed}=2$ model, that 
{decreases to $\sim$34} bar for the  $f_\mathrm{sed}=8$ model at $T_\mathrm{eff}=1000$ K.  The 1.17 $\mu$m $K$ doublet probes lower pressures in the atmosphere than the 1.25 $\mu$m doublet and thus shows a similar trend but of lower amplitude with 
{$\delta P_\mathrm{max}\sim32$ bar} for the $f_\mathrm{sed}=2$ models. Lastly, the 1.14 $\mu$m $Na$ doublet probes comparable pressures with the 1.17 $\mu$m $K$ doublet and thus shows the same trends as the latter. Considering the $K_\mathrm{zz}$ and $f_\mathrm{sed}$ trends (see middle and bottom rows in Fig.~\ref{fig:wide}), both doublets show strong cloud sensitivity for $T_\mathrm{eff}\le1400$~K, with $\delta P_\mathrm{max}\sim6$--8 bar for the $K$ doublet and $\sim5$--7 bar for the $Na$ doublet at $\log g=4.5$. As with the other bands, the largest $\delta P$ occurs for the thickest clouds ($f_\mathrm{sed}=1$) and moderate mixing ($\log K_\mathrm{zz}=8$), and the effect diminishes to $\lesssim2$ bar by $T_\mathrm{eff}=2000$~K. 



\cite{Allers2013} showed that $Na$ and $K$ features appear to be weaker in low-gravity atmospheres. They also discussed the impact of clouds in low-gravity objects, concluding that clouds generally weaken the alkali features,  consistent with other studies of low-gravity and/or dusty brown dwarfs
\citep{Gorlova2003,McGovern2004,Manjavacas2014,Martin2017}. \cite{Morley2024} showed that 
decreasing $f_\mathrm{sed}$ (thicker clouds) in lower $\log g$ atmospheres results in weaker $Na$ and $K$ features.
Our results agree with the work of \cite{Allers2013} that $Na$ and $K$ features appear weaker in low-gravity atmospheres, as well as the \cite{Morley2024} observations that decreasing $f_\mathrm{sed}$ in lower $\log g$ atmospheres results in weaker $Na$ and $K$ features. However, we note that \cite{Morley2024} found that when keeping $f_\mathrm{sed}$ constant, models with lower $\log g$ exhibited stronger $Na$ and $K$ features while we find that increasing  $\log g$ produces stronger $Na$ and $K$ features. They suggest that, in observed brown dwarf spectra, the enhanced cloud thickness typical of low-gravity objects suppresses these spectral features, and this cloud-driven weakening dominates over the opposing influence expected from surface gravity changes alone.
The difference between our results and those of \cite{Morley2024} likely stems from our treatment of vertical mixing. While we maintain a constant $K_\mathrm{zz}$ across our atmospheric profiles, \cite{Morley2024} employ an altitude-dependent $K_\mathrm{zz}$ calculated self-consistently from mixing-length theory (see Sect.~\ref{sec:discussions} \ for further discussion of how constant $K_\mathrm{zz}$ affects cloud formation in our models). 

\section{Discussion}
\label{sec:discussions}

\subsection{Applications of the Grid}

Our contribution–function grid is intended as a practical tool for planning and interpreting brown dwarf observations. To facilitate community access to our contribution function grid, we developed an interactive web application \footnote{\url{https://ltcfgrid.research.ucf.edu/}}. Here we present the tool, and discuss an application on planning and interpreting brown dwarf observations.



\subsection{The Tool } \label{sec:tool}

\begin{figure}[h!]
    \centering
   \includegraphics[trim=13 440 160 53, clip, width=0.68\textwidth]{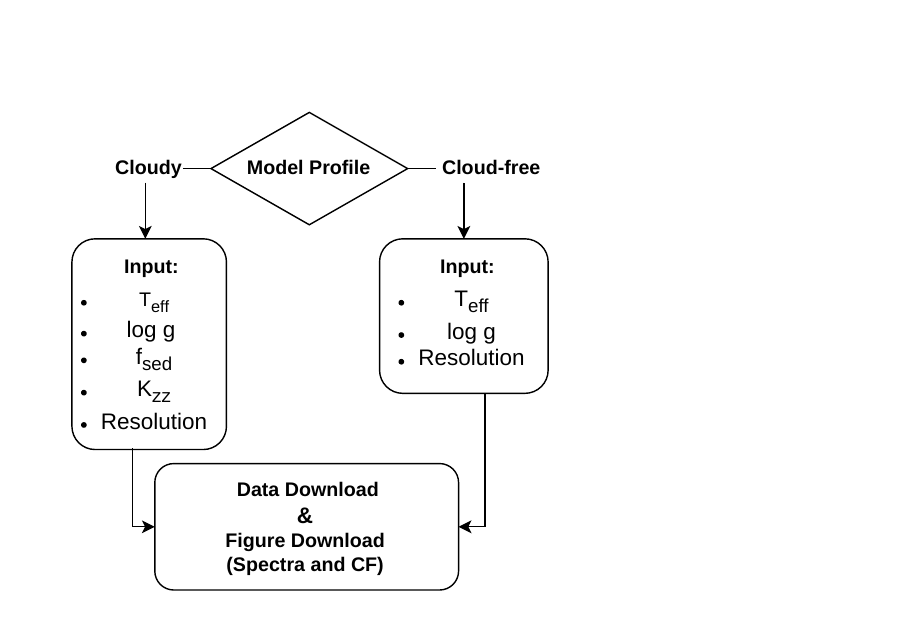}
    \caption{ The tool allows users to input specified grid parameters, allowing for visual synchronous spectral \& \ECF comparison. The model profile output \ECF (CF) data selected is then available for individual download.}
  \label{fig:ToolFig}
\end{figure}
 \begin{figure}[htbp!]
  \centering
  \includegraphics[width= .47\textwidth]{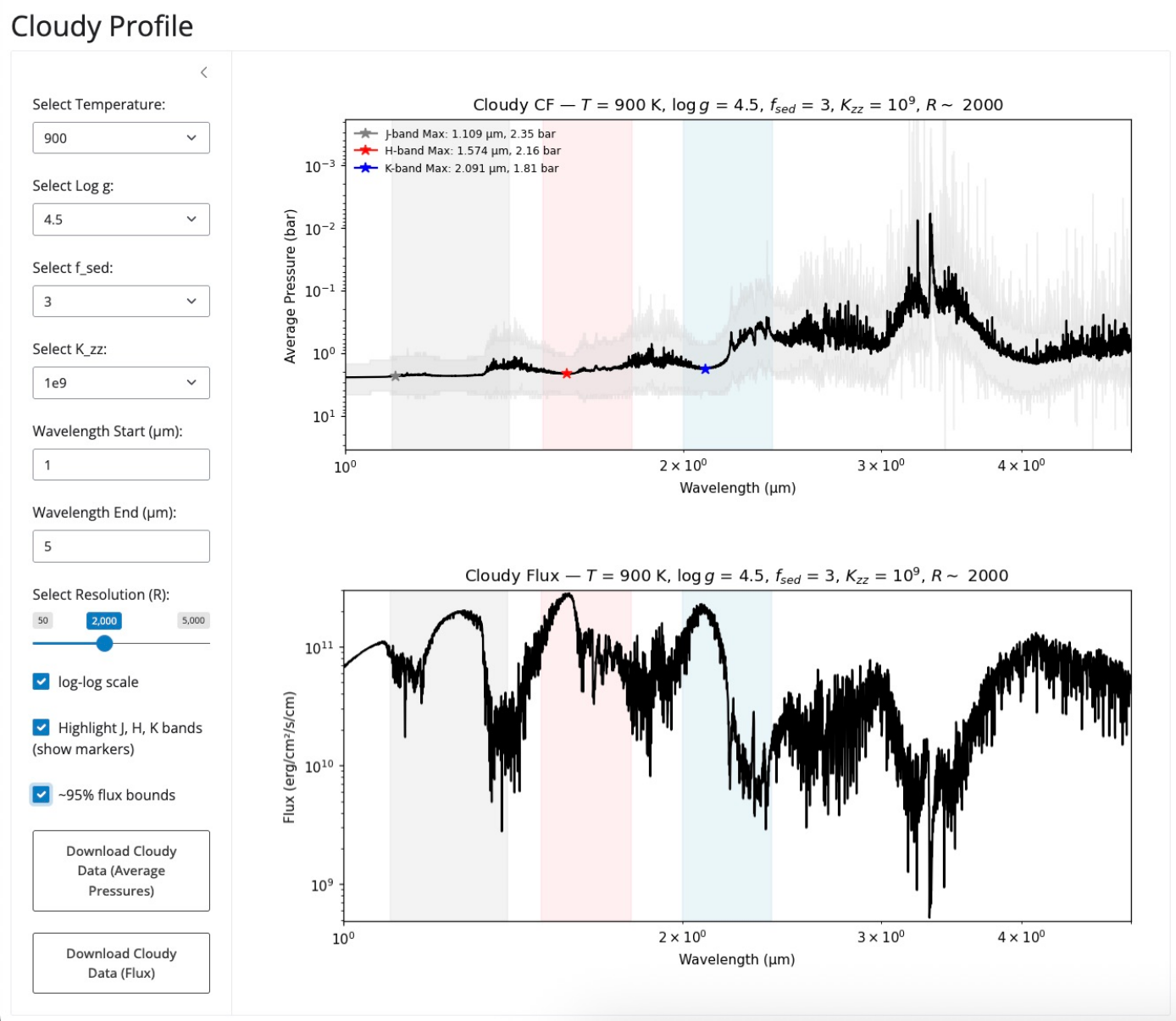}
  \caption{ The tool allows users to input specified grid parameters, allowing for visual synchronous spectral \& \ECF comparison. The model profile output \ECF (CF) data selected is then available for individual download.}
  \label{fig:gui}
\end{figure}


Our GUI allows users to input specific atmospheric parameters ($T_{\text{eff}}$, $\log g$, $f_{\text{sed}}$, $K_\mathrm{zz}$) from our complete parameter space and generates comparisons of model spectra and  contribution functions (Fig.~\ref{fig:ToolFig}). The dual-panel interface (see Fig.~\ref{fig:gui} for a screenshot of the GUI layout) displays both spectra and contribution functions simultaneously, enabling users to directly correlate spectral features with the atmospheric pressure levels they probe. Users can download individual model outputs (at $R \lesssim 5{,}000$) of wavelength-specific contribution functions and spectra as plain ASCII files. Figures can be saved directly by users with a simple drag-and-drop feature or regenerated with customization using downloadable files (see Fig.~\ref{fig:gui}).


\subsubsection{Application of the tool}

Our tool aims to support the characterization of observations and observational planning by identifying the average pressures probed in key wavelengths and the optimal wavelength ranges needed to probe specific pressure levels. To showcase how our tool can be used, we discuss two study cases:

As a first case, we use an observer interested in the vertical cloud structure of a variable L/T transition target. The observer would use the mean observed spectrum of the target and information about the instrument they used (wavelength range, spectral resolution etc) to fit against our grid and identify the best-fit model parameters ($T_\mathrm{eff}$, $\log g$, $f_\mathrm{sed}$, $K_\mathrm{zz}$). Note that the spectral fitting itself would be performed externally by the observer using downloaded model spectra from the tool. Prior knowledge of the target's  $T_\mathrm{eff}$ and $\log g$ can inform this stage to speed up the fit. The observer would then use the corresponding contribution function to identify which pressure ranges are driving the observed light curve variability at different wavelength bands. The observer can thus check the vertical cloud structure of their target, constraining the variability due to, e.g., deeper cloud decks versus upper-atmosphere ones, in a comparable way to what is done in the literature using one-off contribution functions \citep{Chen2024, Chen2025, McCarthy2025, Biller2024, oliverosgomez2025}.

As a second case, we use an observer interested in proposing observations of a target with known $T_\mathrm{eff}$ and $\log g$. We note that  fundamental properties like mass and radius are rarely directly measured for brown dwarfs \citep[e.g.,][]{Li2023,Maire2024,Redai2026}. Instead, both $T_{\rm eff}$ and $\log g$ are inferred from evolutionary models, forward model fitting, or atmospheric retrievals rather than from direct mass and radius measurements \citep[e.g.,][]{Filippazzo2015, Vos_2023}. Prior characterization efforts can provide estimates of these parameters and  guide observational planning. In this case the observer could use our tool to obtain the contribution functions of models within a range of the $T_{\rm eff}$ and $\log g$ of the target (e.g., $T_{\rm eff}\pm$100 K and $\log g\pm$0.5 respectively). They would then convolve the contribution functions with the properties of the instrument they are interested in using, to get the pressures probed with different instrument setups (e.g., \textit{JWST} NIRSpec PRISM/CLEAR vs combining NIRSpec G140M and G395M). The observer would thus be able to plan a better-informed observational campaign, having the ability to test the benefit that different setups or additional filters offer in the study of their target atmosphere. They could thus justify the inclusion (or exclusion) of filters due to their benefit being (or not) justified in comparison to the extra telescope time needed.

\subsection{Comparison to Observational Data}

As a proof-of-concept we compared our grid models with observational data. To quantify the goodness of fit, we used the reduced chi-squared statistic ($\chi^2_\nu$).
First, we applied the \texttt{mean\_regrid} function from \texttt{picaso} to rebin the model spectra to each dataset resolution.
For example, for \textit{JWST}, \textit{NIRSpec}, PRISM/CLEAR BOTS mode observations we rebinned the model spectra to $R\sim$30-300 to produce the final model spectrum. Then, we normalized the model and observed spectra and compared our models against the observation spectra. Finally, we performed the $\chi^2_\nu$ analysis to assess the best-fit model to the observations.


\subsubsection{Application of our grid on HST observations}

\begin{figure*}[htbp!]
    \centering
    \includegraphics[width=1\textwidth]{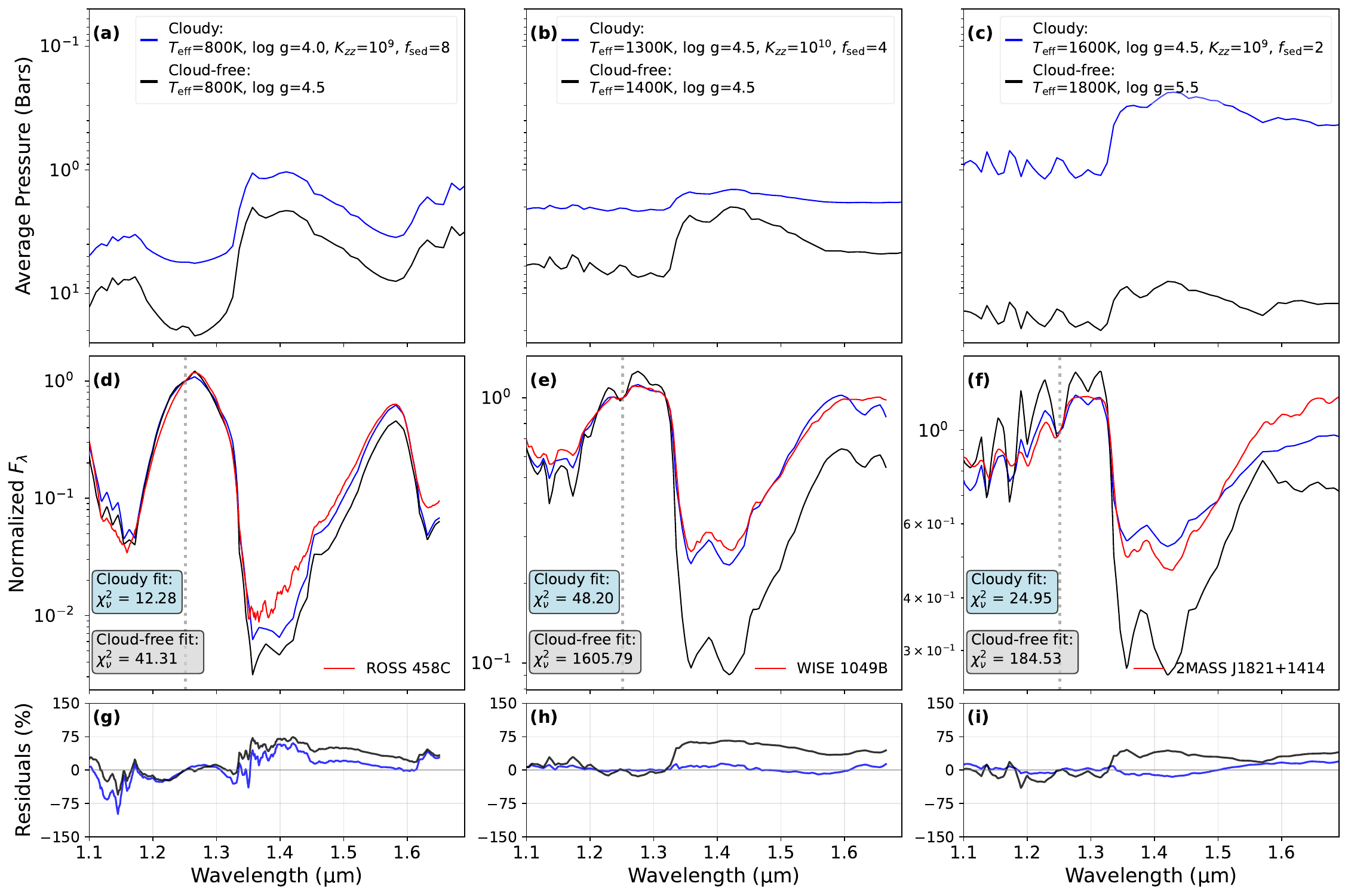}
    \caption{ Contribution functions of our best-fit cloudy and clear models (top panel), observed and best-fit model spectra (middle panels) and their residuals (bottom panels) for Ross 458C (left column), WISE1049B (middle column) and 2M1821 (right column). The observed spectra (red lines) were acquired with \textit{HST/WFC3/G141}. The vertical dotted line shows where the spectra were normalized, at $\lambda \sim 1.25\mu$m. }
  \label{fig:Figobs}
\end{figure*}

We compared our models against three \textit{HST} spectra from the Cloud Atlas spectral library  \citep{Manjavacas_2019}. In  Fig.~\ref{fig:Figobs} (red lines in middle panels) we show the \textit{HST} Wide Field Camera 3/ G141 grism spectra of Ross 458C a late T-type brown dwarf (left column), WISE1049B a L/T transition object (middle column) and 2MASS J1821 a L5 brown dwarf (right column). In the top panels, we show the respective contribution functions for each. We compared these observations against both clear and cloudy models, even though observational studies suggest that these objects are predominantly cloudy \citep{manjavacas2019b,Crossfield2014, Buenzli2014,Yang_2016}. Our best-fit models are consistent with these observations, favoring cloudy atmospheres. We discuss these three objects below.


\underline{Ross 458C:} Earlier studies show that the NIR spectrum of Ross 458C, a late T-type object \citep[T7–9;][]{Morley2012}, is better reproduced by cloudy models and shows variability and signatures consistent with sulfide clouds and modest vertical mixing, as expected for a cool, low-gravity late-T object \citep{Burgasser2010ApJ,Burningham2011MNRAS,Manjavacas2019Ross,Gaarn2023MNRAS}. Comparing the \textit{HST} spectrum of Ross 458C against our grid we found a best-fit with a cloudy model spectrum with $T_\mathrm{eff}=800$K, $\log g=$4, $f_\mathrm{sed}$ = 8 and $\log K_\mathrm{zz}=9$ 
($\chi^2_\nu=$ 12.28; blue line in middle row of Fig.~\ref{fig:Figobs}). Our best-fit cloudy model has the same gravity as that of \citet{Morley2012}, however, we find a best-fit with a hotter $T_\mathrm{eff}$ and thinner clouds ($T_\mathrm{eff}=800$K, $f_\mathrm{sed}$=8 vs $T_\mathrm{eff}=700$K, $f_\mathrm{sed}$=3 of \citet{Morley2012}. Also shown is the best-fit clear atmosphere model spectrum ({$\chi^2_\nu=$ 41.31}; black line in middle row of Fig.~\ref{fig:Figobs}). 
For Ross 458C, our cloudless model is consistent with the best-fit model of \cite{Morley2012} with a $T_{\rm eff}=800$~K, however our model has a higher gravity at $\log g=$4.5 (vs $\log g=$4 of \citet{Morley2012}). The difference between our best-fit models can be attributed to the different wavelength range of our data in comparison to \citet{Morley2012}, with our observations covering only the 1.1$\mu$m-1.7$\mu$m window and that of \citet{Morley2012} an extended 0.8$\mu$m-15$\mu$m window. For example, \citet{Phillips2024} fit medium-resolution ($R\approx1700$) Gemini/GNIRS spectra of Ross~458C with an extended ATMO~2020 grid and found that the inferred parameters depend strongly on the fitted wavelength range. Their full-spectrum ($0.8$--$2.5\,\mu$m) yields a best-fit of $T_{\rm eff}=740\pm50$K and $\log g=3.9\pm0.25$, 
whereas fitting the $Y-$, $J-$, $H-$ and $K-$bands individually returns differing solutions with $T_{\rm eff}\in [700,860]$K and  $\log g\in[3.1,5.4]$, highlighting the dependence of the best-fit model on the wavelength range of the observations. 
Lastly, we note that in the $J-$band the clear and cloudy model spectra for Ross 458C are matching well, although at those wavelengths the corresponding contribution functions (top panels in Fig.~\ref{fig:Figobs}) suggest that the pressures probed in these two models are considerably different (21.8 bar for the cloud-free vs 5.73 bar for the cloudy profile).
This behavior is consistent with a thin, vertically confined condensate cloud layer at low pressure that elevates the apparent photosphere while leaving the $J-$band shape similar to a clear model.

\underline{WISE1049B:} An L/T transition type object \citep[T0.5][]{Burgasser_2013}, is a known variable target whose variability is considered to be driven by clouds \citep[e.g.,][]{Crossfield2014,buenzli_2015,karalidi2016}, or combinations of clouds and thermochemical instabilities at different pressures \citep{Biller2024,Chen2024,oliverosgomez2025}. 
Comparing  the \textit{HST} spectrum of WISE1049B against our grid we found a best-fit with a cloudy model spectrum with $T_\mathrm{eff}=1300$~K, $\log g = 4.5$, $f_\mathrm{sed} = 4$ and $\log K_\mathrm{zz}=10$ ($\chi^2_\nu=$  48.20; blue line in middle row of Fig.~\ref{fig:Figobs}). Also shown is the best-fit clear atmosphere model spectrum ($\chi^2_\nu=$  1605.79; black line in middle row of Fig.~\ref{fig:Figobs}).  Using \textit{HST} observations \citet{buenzli_2015} derived $T_\mathrm{eff}=1300$~K and $\log g = 4.5$, in agreement with our best-fit parameters. Studies using observations at different wavelength ranges and/or  different resolutions, reported similar effective temperatures but favored higher surface gravities:    \citet{Biller2024} using \textit{JWST} NIRSpec PRISM/CLEAR and MIRI observations found a best-fit $T_\mathrm{eff} \sim 1150$ - $1300$~K and $\log g = 4.7 - 5.0 $, while \citet{deregt2025} used high resolution $J-$band \textit{CRIRES+} observations and found a best-fit $\log g = 4.88\pm0.09$. {This further highlights the dependency of the retrieved gravity on the wavelength range observed.}



\underline{2MASS J1821+1414:} (hereafter 2M1821) An L-type object with signs of low-gravity \citep[L4][]{Gagne2015}. The variability of 2M1821 is considered to be driven by clouds and hazes \citep{Yang2015, Yang_2016}. Comparing the \textit{HST} spectrum of 2M1821 against our grid we found a best-fit with a cloudy model spectrum  with $T_\mathrm{eff}=1600$ K, $\log g=$4.5, $f_\mathrm{sed}$ = 2 and $\log K_\mathrm{zz}=9$ ($\chi^2_\nu=$ 24.95; blue line in middle row of Fig.~\ref{fig:Figobs}). Also shown is the best-fit clear atmosphere model spectrum ($\chi^2_\nu=$  184.53; black line in middle row of Fig.~\ref{fig:Figobs}). Our best-fit cloudy models are in agreement with past observations that reported a $T_{\rm eff} = 1635\pm66$~K \citep{Faherty2016}, or $T_{\rm eff}=1600\pm100$~K and $\log g=4.5\pm0.5$ \citep{Gagne2015}.

Overall, our cloudy models have a better fit than the clear atmosphere models for all objects as the cloud opacity strongly dictates the observed thermal emission in the NIR.
In the case of WISE1049B and 2M1821, the addition of clouds smoothed out the contribution function and shifted the continuum to higher altitudes, 
{resulting in a shallower $H_2O$ feature in comparison to the clear atmosphere models} (see \cite{Chen2024} for a discussion of pressure-dependent features on WISE1049B).

\subsection{Application of our grid on JWST Observations of SIMP J013656.5+093347 }

\begin{figure*}[htbp!]
    \centering
    \includegraphics[width=1\textwidth]{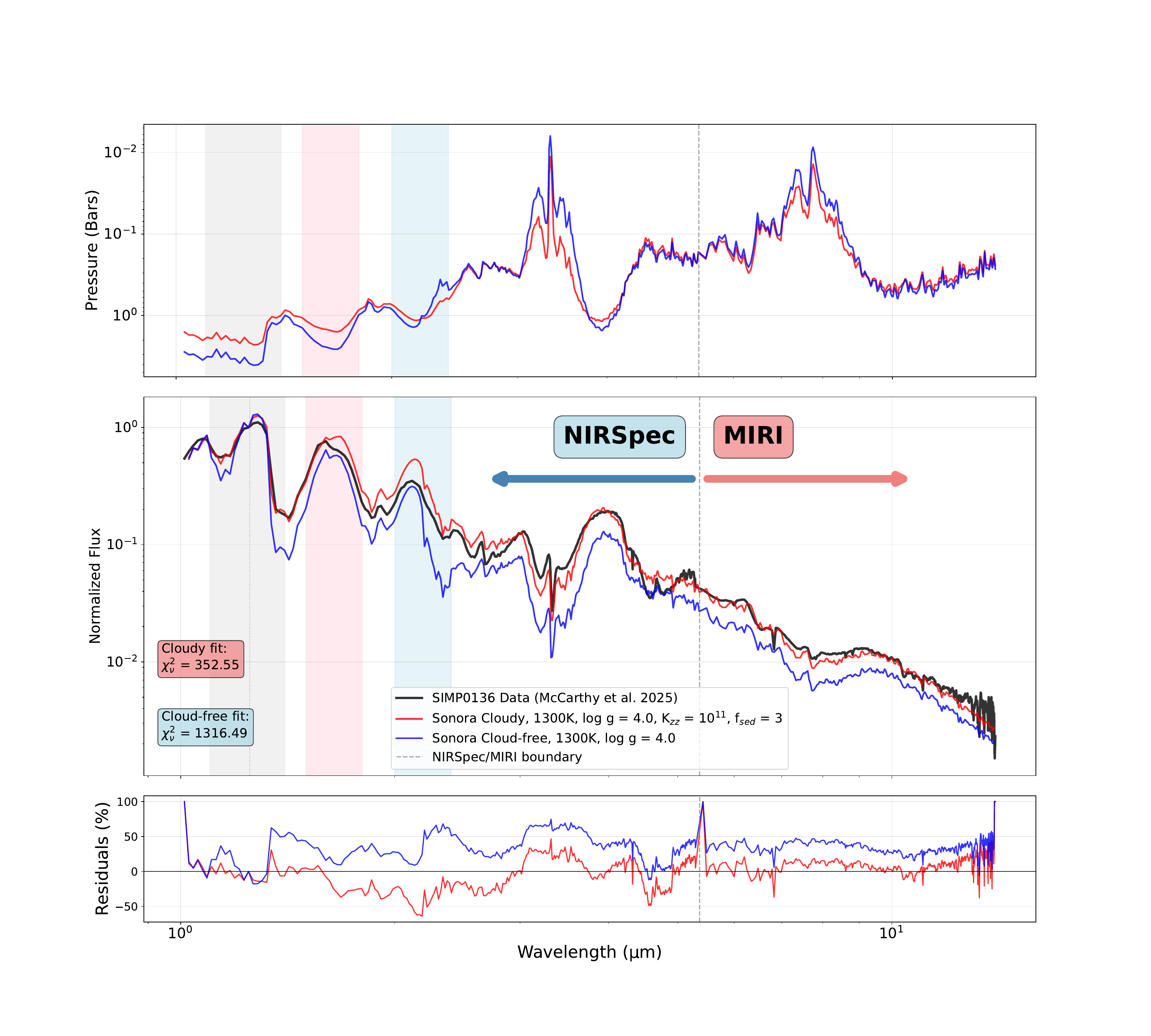}
    \caption{ Similar to Fig.~\ref{fig:Figobs} but for \textit{JWST} {NIRSpec} and {MIRI} observations of SIMP0136 \citep{McCarthy2025}. The dashed vertical line marks the boundary between the NIRSpec and MIRI wavelength ranges. The spectra were normalized at $\sim 1.25 $ $\mu$m.
    }
  \label{fig:FigobsJWST}
\end{figure*}

SIMP J013656.5+093347 (hereafter SIMP0136) is a T2.5 type brown dwarf that is known to show high amplitude variability that can be  attributed to patchy, rotating cloud structures \citep[e.g.,][]{Artigau2009, Apai2013}.
\textit{JWST} ushered in a deeper understanding of atmospheric properties, with several recent studies shedding further light on the complex variability of brown dwarf atmospheres \citep[e.g.,][]{Biller2024, McCarthy2024}. \textit{JWST} observations of SIMP0136 reported a  $T_\mathrm{eff}\sim$1100 K \citep{McCarthy2025},  or $T_\mathrm{eff}=1150\pm70$ K and $\log 4.5\pm 0.4$ based on a spectral energy distribution (SED) analysis \citep{Vos_2023} and  $T_\mathrm{eff}=1329\pm17$ K  and $\log g\  4.25\pm 0.08$  based on a retrieval \citep{Vos_2023}. 
In Fig.~\ref{fig:FigobsJWST} (middle panel) we show the median NIRSPec Bright Object Time Series (BOTS; covers 0.6-5.3 $\mu$m at $R\sim 30$-$300$; \citealp{Jakobsen_NIRSPEC}) and MIRI Low Resolution Spectroscopy (LRS; covers 5-14 $\mu$m with a resolution of $R\sim 40$-$160$; \citealp{Wright2015_MIRI}) Time Series Observations (TSOs) from \cite{McCarthy2025} (black lines).
Comparing the spectrum of SIMP0136 against our grid we found a best-fit with a  cloudy atmosphere $T_\mathrm{eff}=1300$ K, $\log g=4.0$, $f_\mathrm{sed}=3$ and $\log K_\mathrm{zz}=11$ ($\chi^2_\nu=$  352.55; red line in middle panel of Fig.~\ref{fig:FigobsJWST}), in agreement with  the \cite{Vos_2023} retrieval results. We note that the best-fit clear atmosphere model ($\chi^2_\nu=$ 1316.49; blue line in middle panel of Fig.~\ref{fig:FigobsJWST}) also has $T_\mathrm{eff}=1300$ K, $\log g=4.0$.

\subsection{Cloud formation vs $K_\mathrm{zz}$ in our grid}
\label{Cloud Thinning}
\label{subsec: Cloud Thinning }

Through our grid we identified 197 models 
where different $f_\mathrm{sed}-K_\mathrm{zz}$ combinations resulted in identical pressures probed across the $J$-, $H$-, and $K$-bands. All of these models had 
$\log K_\mathrm{zz}\gtrsim11$ and $T_\mathrm{eff}<900$ K. A large number of these models showed the same spectrum as the corresponding clear-atmosphere models, suggesting that at this high $K_\mathrm{zz}$ the vertical mixing was too vigorous and did not allow for the formation of clouds. We note that the 900 K threshold coincides with the threshold that \cite{Morley2024} mentioned, and below which the TP profiles of Diamondback (cloudy atmospheres) coincided with those of Bobcat (clear atmospheres). At these temperatures, the late-T spectral type atmospheres are expected to be clear or have thin clouds \citep{Geballe2001, Morley2012,miles2020}. These clouds could produce moderate warming of the atmosphere \citep[][]{Charnay2018} that could affect the appearance of the spectra and allow for clouds to form in our model atmospheres. 

Clouds follow the \citet{AckermanMarly2001} model as implemented in 
\texttt{Virga} \citep{batalha2025}. In this model the vertical distribution 
of condensate particles is defined by a balance between turbulent mixing and 
gravitational settling:
\begin{equation}
K_\mathrm{zz} \frac{\partial q_t}{\partial z} - f_{\text{sed}} w_* q_c = 0
\label{eq:cloud_balance}
\end{equation}
where $K_\mathrm{zz}$ is the eddy diffusion coefficient, $q_t$ is the total vapor and condensate mixing ratio , $f_{\text{sed}}$ is the 
sedimentation efficiency parameter, $q_c$ is the condensate mass mixing 
ratio, and $w_*$ is the convective velocity scale defined as:
\begin{equation}
w_* = \left( \frac{|g|\,z_i}{T_\mathrm{ML}}\, F_\mathrm{conv} \right)^{\!1/3}
\label{eq:deardorff}
\end{equation}
where $g$ is the gravitational acceleration, $z_i$ is the depth of the 
convective zone, $T_\mathrm{ML}$ is the mixing length temperature, and 
$F_\mathrm{conv}$ is the convective flux \citep{AckermanMarly2001, 
AckermanMarley2013}. Reshuffling Eq.~\ref{eq:cloud_balance} we get:
\begin{equation}
\frac{K_\mathrm{zz}}{f_{\text{sed}} w_*} = \frac{q_c}{\partial q_t/\partial z}
\label{eq:ratio}
\end{equation}
When $\log K_\mathrm{zz}$ increases to vigorous levels ($\geq 10$), strong 
vertical mixing in the model efficiently transports cloud particles between 
atmospheric layers as $w_*$ varies significantly with altitude. When high 
$K_\mathrm{zz}$ transports particles into the upper, low-$w_*$ regions we 
see two primary effects:

1) \textit{Kinetic removal}: Particles are dispersed across larger vertical 
distances before they can maintain local condensate concentrations, reducing 
their optical depth despite potentially conserving the total condensate 
mass; and

2) \textit{Unfavorable thermodynamic conditions}: Particles transported to 
different pressure-temperature regimes may fail to condense efficiently in 
their new environment.
This transport-induced removal process creates the observed ``cloud-free-like'' spectra because radiation can now escape from deeper atmospheric layers where cloud opacity is reduced. This mechanism explains why increasing $K_\mathrm{zz}$ to vigorous levels yields model spectra that appear nearly cloud-free. We note however, that 
the equilibrium relationship in Equation~\ref{eq:ratio} describes the local balance, but doesn't capture the global effect of redistributing particles across regions with vastly different convective velocities. Moreover, our models here assume chemical equilibrium while atmospheres particularly at larger $K_{\rm zz}$ might also exhibit disequilibrium chemistry. Vertical mixing driven by higher $K_{\rm zz}$ can transport molecules (e.g., H$_2$O, CO) upward from deeper atmospheric layers, changing their abundance relative to equilibrium values \citep[e.g.,][]{zahnle2014methane, Lodders2002}. 
Grids of models for cloud-free and cloudy brown dwarfs with disequilibrium chemistry\citep[e.g.,][]{Mukherjee2024,Karalidi2021_cholla, phillips2020} show how vertical mixing can change the atmospheric TP and chemistry profiles relative to equilibrium and impact the emerging spectra, in agreement with observations \citep[e.g.,][]{Saumon2006, Barman2015, miles2020,leggett2024, luhman2024}. Both particle redistribution and disequilibrium chemistry could thus affect the appearance of our spectra and allow the formation of clouds in the model atmospheres. Lastly, we note that \cite{miles2020} fitted observations of five targets with $T_\mathrm{eff}\in[800,500]$ K and constrained $\log K_\mathrm{zz}$ in these atmospheres to be low ($\lesssim 5.3$), in agreement with the fits of three late T-dwarfs by  \citet{Phillips2024} ($\lesssim 8$), which suggests that our $\log K_\mathrm{zz}$ values for these colder models were probably too large.

\section{Conclusions} 
\label{sec:conclusion}
Leveraging \texttt{picaso}, a 1-D radiative transfer code, \texttt{VIRGA}, a cloud generation code, and the clear atmosphere Sonora Diamondback grid, we created a grid of 2407 cloudy and 80 cloud-free brown dwarf atmospheric spectra and \ECFs which spans $T_{eff}$ from 500 K to 2000 K, surface gravities {$\log g\in[3.5,5.5]$ [cgs], sedimentation efficiencies ($f_\mathrm{sed}$) of 1, 2, 3, 4, and 8, and constant with altitude eddy diffusion coefficients ({$K_\mathrm{zz}$) ranging from 10$^6$ to 10$^{12}$ cm$^2$s$^{-1}$. All our models had a constant, solar metallicity and C/O abundance. After checking our models for physical consistency and removing models that did not meet our criteria (see Sect.~\ref{sec:style}), we ended up with a grid of {1539} cloudy and 80 cloud-free brown dwarf spectra and \ECFs with $K_\mathrm{zz}\in[10^6,10^{11}]$ cm$^2$s$^{-1}$.


We confirm that {cloud opacity profoundly impacts atmospheric emission depth across prominent molecular features} such as $CH{_4}$, $H{_2}O$, $CO$, $NH{_3}$, $CO{_2}$, and atomic features $Na$ and $K$.
Overall, the $J-$band consistently exhibited the largest pressure differences} between cloudy and cloud-free profiles due to it probing the deepest in the model atmospheres. This supports the claim that the $J-$band is the {most sensitive spectral window for diagnosing atmospheric depth changes caused by clouds}.
At mid-$R$ the $K-$doublet ($\sim$1.25 $\mu$m), showed the most significant contribution to these differences}: for cooler, high-gravity brown dwarfs with moderate mixing ($\log K_\mathrm{zz}= 8$) and thick clouds $f_\mathrm{sed}=$2 it reached a remarkable $\delta P\sim$60 bar. Additionally, at mid-$R$ the $Na$ ($\sim$1.14 $\mu$m) doublet also exhibited strong cloud sensitivity in the NIR, with $\delta P$ reaching up to $\sim$22 bars at lower $T_\mathrm{eff}$.

{The effect of clouds on the emission depth are strongest at mid-temperatures, around the L/T transition ($T_\mathrm{eff}=1000$ K-$1400$ K), and high gravities}, diminishing at higher {$f_\mathrm{sed}$}, $K_\mathrm{zz}$, and $T_\mathrm{eff}$. 
As expected, cloudy models fit the spectra of L dwarfs best, while many T dwarfs are best-fit with cloud-free models. For high $T_\mathrm{eff}$, vigorous mixing ($\log K_\mathrm{zz}\geq 10$ ) is necessary for the observed flux to originate from deeper within the atmosphere, enabling the appearance of specific absorption features.


To facilitate observational planning and community engagement, we provide an accessible database and developed a user-friendly application that enables seamless comparison of observational data with our model grid of spectra and emission contribution functions. This tool\footnote{https://ltcfgrid.research.ucf.edu/} offers synchronized visual comparisons and allows users to conveniently download individual model spectra and contribution function outputs for further analysis.


Lastly, we note that the cloudy models of this grid are simplified and do not account for variable $f_\mathrm{sed}$ and/or $K_\mathrm{zz}$ with altitude. Future work can explore the \ECFs of 
models considering variable $K_\mathrm{zz}$  \citep[see, e.g.,][]{Mukherjee2022}, while  \texttt{VIRGA} has recently been updated to include altitude-dependent $f_\mathrm{sed}$ \citep{Rooney2022} which would facilitate an extension of our grid. \cite{Mang_2024} compared variable $f_\mathrm{sed}$ \texttt{VIRGA} models to a complex microphysical model (\texttt{CARMA}) and found that the variable $f_\mathrm{sed}$ modification provides greater flexibility in generating cloud profiles and enables the modeling of spectral features that are influenced by cloud properties across different pressure levels, potentially leading to improved agreement with complex models and observed data over wide wavelength ranges. 
Lastly, the grid can be expanded to different metallicities and C/O ratios. As further developments happen, we will update the tool providing the community with an accessible hub to facilitate brown dwarf observation planning. 

\section*{Acknowledgements}
MP, TK, EM and KMM acknowledge support by NASA's Astrophysics Data Analysis
Program grant 80NSSC23K0478. NOG acknowledges support from the Arthur Davidsen Graduate Student Research Fellowship provided by the Space Telescope Science Institute.
This work used computational resources provided by the Stokes high-performance computing cluster at the University of Central Florida's Advanced Research Computing Center (ARCC). We also acknowledge the use of the Jetstream2 at Indiana University through ACCESS allocation CIS230083: Research Facilitation Support for UCF Researchers, which hosted the Shiny for Python application used to explore the model grid. We thank \citet{McCarthy2025} for making
their JWST spectra of SIMP0136 available, which we use for comparison to our grid models.

\newpage









\clearpage
\appendix
\section{Additional Tables}
This appendix collects supplementary material referenced in the main text. Table~\ref{tab:table_pres} lists the average pressures probed in the $J-$, $H-$, and $K-$ bands for the model atmospheres shown in Figs.~\ref{fig:LTcf_fsed2} and~\ref{fig:PressurePanel_3panel_4.5_Kzz8}, spanning $T_\mathrm{eff}=700$, 1300, and 1900~K, $\log g=3.5$, 4.5, and 5.5, $f_\mathrm{sed}=2$, 4, and 8, and $\log K_\mathrm{zz}=8$, 9, and 10. Table~\ref{tab:teff_spt_mapping} gives the mapping between the $T_\mathrm{eff}$ values annotated with spectral types in Fig.~\ref{fig:heatmap_results_Kzz1e8} and the closest empirically determined $T_\mathrm{eff}$, $\log g$, and spectral type from the \citet{Filippazzo2015} sample.
\begin{table}[H]
    \centering
    \begin{tabular}{cccc}
        \hline
        Model &
        Pressure probed &
        Pressure probed &
        Pressure probed \\
        ($T_\mathrm{eff},f_\mathrm{sed},\log g, \log K_\mathrm{zz}$) &
        in $J-$band (bar) &
        in $H-$band (bar) &
        in $K-$band (bar) \\
        \hline
        700, 2, 3.5, 8   &  0.462  & 0.380  & 0.331  \\
        700, 2, 3.5, 9   &  0.925  & 0.728  & 0.526  \\
        700, 2, 3.5, 10  &  1.479  & 1.225  & 0.790  \\
        1300, 2, 3.5, 8  &  0.055  & 0.049  & 0.043  \\
        1300, 2, 3.5, 9  &  0.101  & 0.0927 & 0.0873 \\
        1300, 2, 3.5, 10 &  0.151  & 0.139  & 0.132  \\
        1900, 2, 3.5, 8  &  0.003  & 0.003  & 0.004  \\
        1900, 2, 3.5, 9  &  0.022  & 0.010  & 0.007  \\
        1900, 2, 3.5, 10 &  0.547  & 0.332  & 0.240  \\
        700, 2, 4.5, 10  &  10.862 & 7.846  & 3.405  \\
        700, 2, 4.5, 9   &  6.07   & 4.91   & 2.74   \\
        1300, 2, 4.5, 10 &  3.623  & 1.253  & 0.733  \\
        1300, 2, 4.5, 9  &  0.332  & 0.282  & 0.261  \\
        1900, 2, 4.5, 10 &  3.501  & 2.378  & 1.700  \\
        1900, 2, 4.5, 9  &  2.95   & 1.65   & 0.98   \\
        700, 2, 4.5, 8   &  2.989  & 2.417  & 1.921  \\
        700, 4, 4.5, 8   &  9.064  & 6.400  & 2.746  \\
        700, 8, 4.5, 8   &  15.374 & 8.646  & 3.231  \\
        1300, 2, 4.5, 8  &  0.193  & 0.174  & 0.161  \\
        1300, 4, 4.5, 8  &  1.102  & 1.065  & 1.019  \\
        1300, 8, 4.5, 8  &  2.093  & 1.971  & 1.732  \\
        1900, 2, 4.5, 8  &  0.034  & 0.020  & 0.018  \\
        1900, 4, 4.5, 8  &  1.644  & 0.635  & 0.322  \\
        1900, 8, 4.5, 8  &  2.990  & 1.597  & 0.957  \\
        700, 2, 5.5, 8   &  23.791 & 18.406 & 9.793  \\
        700, 2, 5.5, 10  &  96.496 & 41.498 & 11.559 \\
        700, 2, 5.5, 9   &  49.3   & 33.9   & 11.2   \\
        1300, 2, 5.5, 8  &  1.754  & 0.886  & 0.627  \\
        1300, 2, 5.5, 10 &  26.168 & 13.799 & 6.679  \\
        1300, 2, 5.5, 9  &  17.7   & 7.81   & 3.54   \\
        1900, 2, 5.5, 8  &  13.043 & 6.658  & 3.065  \\
        1900, 2, 5.5, 10 &  18.000 & 11.878 & 7.463  \\
        1900, 2, 5.5, 9  &  17.1   & 10.8   & 6.36   \\
        \hline
    \end{tabular}
    \caption{Pressures probed in the $J-$, $H-$, and $K-$ bands for our models shown in Figs.~\ref{fig:LTcf_fsed2} and~\ref{fig:PressurePanel_3panel_4.5_Kzz8}}
    \label{tab:table_pres}
\end{table}

\begin{table}[H]
    \centering
    \begin{tabular}{rrrrrcc}
        \hline
        Target $T_{\mathrm{eff}}$ (K) &
        Closest $T_{\mathrm{eff}}$ (K) &
        $\sigma_{T_{\mathrm{eff}}}$ (K) &
        $\log g$ (dex) &
        $\sigma_{\log g}$ (dex) &
        Designation &
        Inferred spectral type \\
        \hline
        500  &  569  &  45  & 4.68 & 0.53 & UGPS J072227.51--054031.2       & T9   \\
        800  &  880  &  76  & 4.96 & 0.49 & 2MASSI J1047538+212423          & T6.5 \\
        1100 & 1228  & 127  & 5.02 & 0.48 & SDSSp J175032.96+175903.9       & T3.5 \\
        1400 & 1334  &  58  & 4.99 & 0.26 & Luhman 16A                      & L8   \\
        1700 & 1659  &  74  & 5.18 & 0.22 & SDSS J053951.99--005902.0       & L5   \\
        2000 & 2044  &  87  & 5.22 & 0.15 & 2MASSI J2057540--025230         & L1.5 \\
        \hline
    \end{tabular}
    \caption{
        Mapping between the subset of target $T_{\mathrm{eff}}$ values 
        (500, 800, 1100, 1400, 1700, and 2000 K) for which we explicitly
        annotate spectral types, and the closest empirically determined
        $T_{\mathrm{eff}}$ from the \cite{Filippazzo2015} sample. The table
        lists the closest $T_{\mathrm{eff}}$, its quoted uncertainty,
        surface gravity $\log g$ and its uncertainty, the object
        designation, and the inferred spectral type used in our figures.
    }
    \label{tab:teff_spt_mapping}
\end{table}





\software{\textsc{picaso} v3.2 \citep[][Zenodo: \citealt{batalha_picaso_zenodo}]{Mukherjee2023}, 
          \textsc{virga} v0.0 \citep[][Zenodo: \citealt{batalha_virga_zenodo}]{batalha2025}}
\bibliographystyle{aasjournal}
\bibliography{Contribution_Function_Grid}



\end{document}